\documentclass[twocolumn,showpacs,showkeys]{revtex4}
\usepackage{graphicx,subfigure}

\newcommand{\bastar}{\begin{eqnarray*}}
\newcommand{\eastar}{\end{eqnarray*}}
\newskip\humongous \humongous=0pt plus 1000pt minus 1000pt

\newif\ifdtup

\relax
\newcommand{\bea}{\begin{eqnarray}}
\newcommand{\eea}{\end{eqnarray}}
\newcommand{\nn}{\nonumber}
\newcommand{\pro}{\partial}
\newcommand{\oneg}{\displaystyle\frac{1}{g}}
\newcommand{\dfrac}{\displaystyle\frac}
\newcommand{\mn}{{\mu\nu}}
\newcommand{\A}{{\vec A}}
\newcommand{\bA}{{\bar A}}

\newcommand{\hA}{{\hat A}}
\newcommand{\B}{{\vec B}}

\newcommand{\hB}{{\hat B}}
\newcommand{\tB}{{\widetilde{B}}}
\newcommand{\C}{{\vec C}}

\newcommand{\tC}{{\widetilde{C}}}
\newcommand{\hD}{{\hat D}}
\newcommand{\bD}{{\bar D}}
\newcommand{\cD}{{\cal D}}

\newcommand{\E}{{\vec E}}
\newcommand{\e}{{\vec e}}
\newcommand{\he}{{\hat e}}

\newcommand{\F}{{\vec F}}
\newcommand{\bF}{{\bar F}}

\newcommand{\hF}{{\hat F}}
\newcommand{\vg}{{\bf g}}
\newcommand{\hvg}{\hat {{\bf g}}}
\newcommand{\vG}{{\bf G}}

\newcommand{\tG}{{\widetilde{G}}}
\newcommand{\Gm}{{\Gamma}}

\newcommand{\vGm}{{\bf \Gamma}}
\newcommand{\hvGm}{\hat{{\bf \Gamma}}}
\newcommand{\tGm}{{\widetilde{\Gamma}}}
\newcommand{\tH}{{\widetilde{H}}}
\newcommand{\vI}{{\bf I}}
\newcommand{\vj}{{\bf {j}}}
\newcommand{\tj}{\tilde{j}}
\newcommand{\vtj}{\tilde{{\bf j}}}
\newcommand{\tJ}{{\widetilde{J}}}
\newcommand{\vk}{{\bf {k}}}
\newcommand{\tk}{\tilde{k}}
\newcommand{\vtk}{\tilde{{\bf k}}}
\newcommand{\tK}{{\widetilde{K}}}
\newcommand{\vl}{{\bf {l}}}
\newcommand{\tl}{\tilde{l}}
\newcommand{\vtl}{\tilde{{\bf l}}}
\newcommand{\tL}{{\widetilde{L}}}
\newcommand{\cJ}{{\cal J}}
\newcommand{\cK}{{\cal K}}
\newcommand{\cL}{{\cal L}}
\newcommand{\ctJ}{\widetilde{{\cal J}}}
\newcommand{\ctK}{\widetilde{{\cal K}}}
\newcommand{\ctL}{\widetilde{{\cal L}}}
\newcommand{\m}{{\vec m}}
\newcommand{\hm}{{\hat m}}
\newcommand{\M}{{\vec M}}
\newcommand{\n}{{\vec n}}

\newcommand{\hn}{{\hat n}}
\newcommand{\vP}{{\bf \Pi}}
\newcommand{\vp}{{\bf p}}
\newcommand{\vtp}{\tilde{\bf p}}
\newcommand{\vR}{{\bf R}}
\newcommand{\hR}{{\hat R}}
\newcommand{\hvR}{\hat {\bf R}}

\newcommand{\Si}{{\Sigma}}
\newcommand{\vSi}{{\bf {\Sigma}}}
\newcommand{\tU}{\widetilde {U}}
\newcommand{\tV}{\widetilde {V}}
\newcommand{\tW}{\widetilde {W}}
\newcommand{\X}{{\vec X}}
\newcommand{\pX}{{\vec X}'}
\newcommand{\dX}{\dot{\vec X}}
\newcommand{\Y}{{\vec Y}}
\newcommand{\pY}{{\vec Y}'}
\newcommand{\dY}{\dot{\vec Y}}
\newcommand{\vZ}{{\bf Z}}
\newcommand{\tZ}{\widetilde{Z}}
\newcommand{\pZ}{{\bf Z}'}
\newcommand{\dZ}{\dot{\bf Z}}
\newcommand{\Za}{{Z}^1}
\newcommand{\Zb}{\widetilde{Z}^1}
\newcommand{\Zc}{{Z}^2}
\newcommand{\Zd}{\widetilde{Z}^2}
\begin{document}
\title {Abelian decomposition of Einstein's theory:\\
Reformulation of general relativity}
\author{Y. M. Cho}
\email{ymcho@yongmin.snu.ac.kr}
\author{Sang-Woo Kim}
\email{eyed@phya.snu.ac.kr}
\author{J. H. Kim}
\email{rtiger@phya.snu.ac.kr}
\affiliation{Center for Theoretical
Physics and School of Physics, College of Natural Sciences, Seoul
National University, Seoul 151-742, Korea  \\}
\begin{abstract}
~~~~~We propose a reformulation of general relativity by
making the Abelian decomposition of Einstein's theory.
Based on the view that Einstein's theory can be
interpreted as a gauge theory of Lorentz group,
we decompose the gravitational connection
(the gauge potential of Lorentz group $\vGm_\mu$) into
the restricted connection made of the potential of
the maximal Abelian subgroup $H$ of Lorentz group $G$ and
the valence connection made of $G/H$ part of the potential which
transforms covariantly under Lorentz group. With this decomposition
we show that the Einstein's theory can be decomposed into
the restricted part made of the restricted connection
which has the full Lorentz gauge invariance and the valence
part made of the valence connection which plays
the role of gravitational source of the restricted gravity.
We show that there are two different Abelian decomposition
of Einstein's theory, the space-like decomposition and 
the light-like decomposition, because Lorentz group has two 
maximal Abelian subgroups. 
In this decomposition the role of the space-time metric
$g_\mn$ is replaced by a Lorentz covariant 
four-index metric tensor $\vg_\mn$
which transforms covariantly under the Lorentz group,
and the metric-compatibility condition $\nabla_\alpha g_\mn=0$
of the connection is replaced by the gauge covariant condition
$D_\mu \vg^\mn=0$ which tells that $\vg_\mn$
is invariant under the parallel transport along
the $\pro_\mu$-direction defined by $\vGm_\mu$.
We discuss the physical implications of the Abelian decomposition.
In particular, we argue that the decomposition implies the existence of
a restricted theory of gravitation which has the full general invariance
but has less physical degrees of freedom.
\end{abstract}
\maketitle

\section{Introduction}

Einstein's theory of gravitation and the gauge theory
(Abelian and non-Abelian) are two
fundamental ingredients of theoretical physics which
have played a crucial role to advance our understanding of nature.
Einstein's theory has been very successful describing
the gravitational force. At the same time it has played a
crucial role in unified theory \cite{kal,jord}. In fact all modern
unified theories, including the superstring, are based on
Einstein's theory one way or another. This originates from the
general invariance principle of Einstein's theory which
guarantees that all physical laws should be independent of
the observers. For this reason Einstein's theory has been
the subject of intensive theoretical study.

The gauge theory, the other important ingredient in theoretical
physics, describes the electroweak and strong forces. But the gauge
theory is closely related to Einstein's theory. Indeed the gauge
theory itself can be viewed as an Einstein's theory originating from
the extrinsic curvature of (4+n)-dimensional unified space. 
It is well-known that the (4+n)-dimensional 
unified space made of the 4-dimensional
space-time and an n-dimensional internal space, the
(4+n)-dimensional Einstein's theory reproduces the gauge theory when
the internal space has an n-dimensional isometry $G$.
In fact the (4+n)-dimensional Einstein's theory provides a natural
unification of gauge theory with gravitation, which is known as the
Kaluza-Klein miracle \cite{kal,jord,jmp75,prd75,plb87}.

Conversely Einstein's theory itself can be understood as a
gauge theory, because the general invariance of Einstein's theory
can be viewed as a gauge invariance. The idea that
the general invariance can be associated to a gauge invariance
has been proposed by many people, but the actual application
of this idea to Einstein's theory has been
non-trivial \cite{kib,uti,car,prd76a,prd76b,hehl}. 
There are two ways to formulate the Einstein's theory as a
gauge theory. One can view it as a gauge theory
of 4-dimensional translation group, because the local
4-dimensional translation can be identified as the general
coordinate transformation \cite{kib}. In this case 
one can identify the gauge potential of the translation group 
as the (non-trivial part of) the tetrad \cite{prd76a}.
Or, one can view it as a gauge theory of Lorentz
group (or Poincare group in general), because
the Lorentz gauge transformation can also be
interpreted as the general coordinate 
transformation \cite{uti,car}. In this case one can identify 
the gauge potential of Lorentz group as the spin 
connection \cite{prd76b,hehl}. This tells that
the two theories are closely related.

During the last few decades our understanding of non-Abelian gauge
theory has been extended very much. By now it has been well known
that the non-Abelian gauge potential allows the Abelian
decomposition \cite{prd80,prl81}. It can be decomposed into the
restricted potential of the maximal Abelian 
subgroup $H$ of the gauge group $G$
which has an electric-magnetic duality and the valence
potential of $G/H$ which transforms covariantly under $G$.
A remarkable feature of this decomposition is that 
it is gauge independent. As importantly, the
restricted potential has the full non-Abelian gauge degrees of
freedom, in particular the topological degrees of the gauge group
$G$, in spite of the fact that it consists of only the Abelian
degrees of the maximal Abelian subgroup $H$. This means
that we can construct a restricted gauge theory, a non-Abelian
gauge theory made of only the restricted potential which has much
less physical degrees of freedom, which nevertheless has
the full gauge invariance. Furthermore, we can recover the
full non-Abelian gauge theory simply by adding the valence part.
This tells that the non-Abelian gauge theory can be interpreted
as a restricted gauge theory which has the valence potential
as the gauge covariant source \cite{prd80,prl81}. The importance of 
this decomposition is that the restricted part
plays a fundamental role for us to establish the Abelian dominance 
in non-Abelian dynamics, which have been pivotal in proving
the monopole condensation and confinement of color 
in QCD \cite{fadd,prd02,jhep04,kondo}.

The purpose of this paper is to present a similar Abelian
decomposition of Einstein's theory, regarding the theory
as a gauge theory of Lorentz group. 
{\it Applying the Abelian decomposition
to the gauge potential of Lorentz group,
we show that we can decompose the gravitational connection
to the restricted connection and the valence connection.
With this we decompose the Einstein's theory into
the restricted part made of the restricted connection
and the valence part made of the gauge covariant
valence connection. The decomposition tells that
the Einstein's theory can be viewed as
a restricted theory of gravitation which has the gauge covariant
valence connection as the gravitational source. We show that
Einstein's theory allows two different Abelian decompositions, 
the space-like decomposition and the light-like decomposition,
because the Lorentz group has two maximal Abelian subgroups.}
To decompose the Einstein's theory, we introduce the concept of
the gauge covariant metric $\vg_\mn$, a four-index tensor
$g_\mn^{~~ab}$ which forms an adjoint representation of
Lorentz group, and show that the metric-compatibility
condition of the gravitational connection $\nabla_\alpha g_\mn=0$
is transformed to the gauge covariant condition $D_\mu \vg^\mn=0$
which assures the invariance of $\vg_\mn$ under
the parallel transport along the $\pro_\mu$-direction.

Of course, Einstein's theory as a gauge theory of Lorentz group
is different from the ordinary non-Abelian gauge theory \cite{prd75}.
In gauge theory the fundamental field is the gauge potential,
but in Einstein's theory the fundamental field is the metric.
And in the gauge formulation the gauge potential of Lorentz group
corresponds to the gravitational connection, not the metric.
Also, in gauge theory the Yang-Mills Lagrangian is quadratic in
field strength. But in gravitation the
Einstein-Hilbert Lagrangian is made of the scalar curvature,
which is linear in field strength \cite{prd76b}.
Nevertheless we can still make the Abelian decomposition of the
gravitational connection, and express the Einstein-Hilbert Lagrangian
in terms of the restricted connection and the valence connection.
With this we can separate the restricted part of gravitation
from the Einstein's theory, and show that the theory
can be interpreted as a restricted theory of gravity
which has the valence connection as the gravitational source.

The paper is organized as follows. In Section II we review
the geometry of gauge theory and Abelian decomposition
for later purpose. In Section III
we discuss the simplest Abelian decomposition, the $U(1)$
decomposition of $SU(2)$ gauge theory, as an example to help
us to understand the Abelian decomposition of
Einstein's theory. In Section IV we show how to decompose
the gravitational connection to the Abelian part and the
valence part. We show that there are two different ways
of Abelian decomposition, because the Lorentz group
has two maximal Abelian subgroups. In Section V introduce
the concept of the Lorentz covariant metric tensor, and show
how to decompose the Einstein's theory to the restricted
part and the valence part. We discuss two different
decompositions of Einstein's theory separately.
Finally in Section VI we discuss
the physical implications of our results.

\section{Geometry of Gauge Theory: A Review}

To understand how to make the desired Abelian decomposition of
Einstein's theory, we have to understand the Abelian decomposition
of gauge theory first. So in this section we review 
the geometry of gauge theory and Abelian decomposition. 
Let $P$ be a (4+n)-dimensional unified space endowed with the
metric $g_{AB}~(A,B=1,2,...4+n)$, which has an n-dimensional
linearly independent isometry (the left isometry) which forms
an isometry group $G$. Let the isometry
be described by $n$ linearly independent vector fields $\xi_i$,
\bea
&[\xi_i,~\xi_j]=\dfrac{1}{\kappa} f_{ij}^{~~k} \xi_k,~~~(i,j,k=1,2,...,n) \nn\\
&{\cal L}_{\xi_i} ~g_{AB} =0,~~~(A,B=1,2,...4+n)
\eea
where $\kappa$ is the scale parameter which determines
the length of the isometry vector fields, and ${\cal L}_{\xi_i}$
is the Lie derivative along $\xi_i$.
Clearly the integral manifold of the isometry vector fields forms an
n-dimensional group manifold $G$ which acts vertically on $P$. Let
$\Pi$ be the horizontal projection of $p \in P$ to $x \in M=P/G$,
and identify the 4-dimensional quotient space $M$ as the space-time
manifold. With this we can view the unified space as a
(4+n)-dimensional fiber bundle $P(M,G)$ made of the 4-dimensional
space-time $M$ as the base manifold and the n-dimensional group
space $G$ as the fiber space on which $G$
acts as the structural group \cite{jmp75,prd75}.

Let $\pro_\mu~(\mu=0,1,2,3)$ be a coordinate basis on $M$,
\bea
[\pro_\mu,~\pro_\nu]=0.
\eea
and $D_\mu=\Pi^{-1} \cdot \pro_\mu$ be the horizontal lift
of $\pro_\mu$ on $P$. Clearly $D_\mu$ and $\xi_i$ forms a
basis of $P$ which has the following commutation relations
\cite{jmp75,prd75},
\bea
&[D_\mu,~D_\nu]=-\kappa F_\mn^{~~k}\xi_k,  \nn\\
&[D_\mu,~\xi_i]=0,
~~~[\xi_i,~\xi_j]=\dfrac{1}{\kappa} f_{ij}^{~~k} \xi_k.
\label{hlb}
\eea
Moreover, since $D_\mu$ are horizontal, we have
\bea
&g_{AB} D_\mu^{~A} D_\nu^{~B}= \gamma_\mn,  \nn\\
&g_{AB} D_\mu^{~A} \xi_i^{~B} =0, \nn\\
&g_{AB} \xi_i^{~A} \xi_j^{~B}=\phi_{ij}.
\eea
So, in the basis (\ref{hlb})
the metric acquires the following form
\bea
&g_{AB}=\left( \begin{array}{cc} \gamma_\mn &0\\0& \phi_{ij}
\end{array} \right).
\label{fbm1}
\eea
This suggests us to identify $g_\mn$ and $\phi_{ij}$ as 
the 4-dimensional metric on the space-time manifold $M$ 
and the n-dimensional internal metric on the fiber space $G$. 
From (\ref{hlb}) and (\ref{fbm1}) we can calculate
the Einstein-Hilbert action on $P$. With
\bea
&g= {\rm Det}~(g_\mn),~~~~\phi={\rm Det}~(\phi_{ij}), \nn\\
&\phi_{ij}=\phi^{1/n}~\rho_{ij},~~~~|{\rm Det}~\rho_{ij}|=1,
\eea
we have (up to a total divergence) \cite{jmp75,prd75}
\bea
&S_P=-\dfrac{V_G}{16\pi G_P} \int {\sqrt -\gamma}
{\sqrt \phi} \Big\{R_M +R_G \nn\\
&-\dfrac{\kappa^2}4 \phi^{1/n} \rho_{ij} \gamma^{\mu\rho} \gamma^{\nu\sigma}
F_\mn^{~~i} F_{\rho\sigma}^{~~j} \nn\\
&+\dfrac14 \rho^{ij} \rho^{kl} 
\gamma^\mn (D_\mu \rho_{ik})(D_\nu \rho_{jl}) 
-\dfrac{n-1}{4n} \dfrac{\gamma^\mn 
(\pro_\mu \phi)(\pro_\nu \phi)}{\phi^2} \nn\\
&-\Lambda_P +\lambda(|Det \rho_{ij}|-1) \Big\}~d^4 x, 
\label{cact}
\eea
where $G_P$ and $\Lambda_P$ are the (4+n)-dimensional
gravitational constant and cosmological constant of $P$,
$V_G$ is the n-dimensional volume
of the group space $G$, $R_M$ and $R_G$ are the scalar curvatures 
of the 4-dimensional metric $\gamma_\mn$ and the n-dimensional
metric $\phi_{ij}$.

To clarify the meaning of the unified action (\ref{cact}) 
we introduce another basis defined by a cross section
$\sigma(x)$ in $P$. Let $U$ be an open subset of $M$, and
$\sigma(x)$ be a 4-dimensional submanifold in $P$ which is
diffeomorphic to $U$ with $\Pi \cdot \sigma(x)=x \in U$. Clearly we can
view $\sigma$ as a mapping from $U$ to $\sigma \subset P$,
\bea
&\sigma:~~~x \in U \rightarrow \sigma(x) \in \Pi^{-1}(U).
\eea
Let $\tilde \pro_\mu^{~(\sigma)}=\sigma \cdot \pro_\mu$ 
be the mapping of the vector field $\pro_\mu$ on $\sigma(x)$ 
induced by $\Pi^{-1}$, and let $\tilde \pro_\mu$ be 
the vector field induced on $\Pi^{-1}(U)$ obtained by 
the vertical translation of $\tilde \pro_\mu^{~(\sigma)}$.
Clearly $\tilde \pro_\mu$ and $\xi_i$ form a basis
on $\Pi^{-1}(U)$ which has the following commutation relations
\bea
&[\tilde \pro_\mu,~\tilde \pro_\nu]=0,  \nn\\
&[\tilde \pro_\mu,~\xi_i]=0,
~~~[\xi_i,~\xi_j]=\dfrac{1}{\kappa} f_{ij}^{~~k} \xi_k.
\label{cb}
\eea
Let
\bea
&D_\mu=\tilde \pro_\mu-\kappa A_\mu^{~i} \xi_i, \nn\\
&g_{AB} \xi_i^{~A} \tilde \pro_\mu^{~B} =\kappa \phi_{ij} A_\mu^{~j},
\label{cbmu}
\eea
and find that in this basis the metric is written as
\bea
&g_{AB}=\left( \begin{array}{cc} g_\mn & \kappa A_\mu^{~i} \phi_{ij}
\\ \kappa \phi_{ij} A_\nu^{~j} & \phi_{ij}
\end{array} \right). 
\label{fbm2}
\eea
Moreover, with (\ref{cb}) and (\ref{cbmu}) we have
\bea
&\pro_{\xi_i} A_\mu^{~k}=-\dfrac{1}{\kappa} f_{ij}^{~~k} A_\mu^{~j}, \nn\\
&F_\mn^{~~k}=\pro_\mu A_\nu^{~k}-\pro_\nu A_\mu^{~k}
+ f_{ij}^{~~k} A_\mu^{~i} A_\nu^{~j}.
\eea
This tells that $A_\mu^{~k}$ and $F_\mn^{~~k}$ can be interpreted
as the gauge potential and the field strength of
the isometry group $G$ (in which the coupling
constant $g$ is normalized to be the unit), and
the horizontal lift basis $D_\mu$ plays the
role of the gauge covariant derivative.

Now, with
\bea
&g_\mn= \sqrt \phi~\gamma_\mn,
~~~\varphi=\dfrac12 \sqrt{\dfrac{n+2}{n}}~ \ln \phi,
\eea
(\ref{cact}) is written as
\bea
&S_P=-\dfrac{\sqrt -g}{16\pi G_N} \int \Big\{R_4 
+\dfrac12 g^\mn (\pro_\mu \varphi)(\pro_\nu \varphi)  \nn\\
&+\hat R_G \exp \big(-\dfrac{n+2}{n}~\varphi \big)
-\Lambda_P \exp \big(-\dfrac{n}{n+2}~\varphi \big) \nn\\
&-\dfrac{\kappa^2}4 \exp \big(\dfrac{n+2}{n}~\varphi \big)\rho_{ij} 
g^{\mu\rho} g^{\nu\sigma} F_\mn^{~~i} F_{\rho\sigma}^{~~j} \nn\\
&+\dfrac14 \rho^{ij} \rho^{kl} g^\mn (D_\mu \rho_{ik})(D_\nu \rho_{jl})  \nn\\
&+\lambda(|Det \rho_{ij}|-1) \Big\}~d^4 x,
\label{cact}
\eea
where $\hat R_G$ is the scalar curvature of the normalised metric
$\rho_{ab}$,
\bea
&\hat R_G(\rho_{ab})=-\dfrac12 f_{ab}^{~~d}f_{cd}^{~~b} \rho^{ac}
-\dfrac 14 f_{ab}^{~~m}f_{cd}^{~~n}\rho^{ac}\rho^{bd}\rho_{mn}, \nn
\eea
This tells that the (4+n)-dimensional Einstein's
theory indeed describes the unified theory of the 4-dimensional
Einstein's theory and the gauge theory of the isometry group $G$ 
coupled to the gauge covariant internal metric
$\phi_{ij}$, provided that \cite{jmp75,prd75}
\bea
&\dfrac{V_G}{16\pi G_P}=\dfrac{1}{16\pi G_N},
~~~~\dfrac{\kappa^2}{16\pi G_N}=1,
\label{ccond}
\eea
where $G_N$ is the 4-dimensional Newton's gravitational constant.
In this unification the isometry group $G$ plays the role
of the gauge group, and the choice of a cross section
corresponds to the choice of a gauge \cite{jmp75,prd75}.
Notice, however, that the metric $g_\mn$ on $M$ is the Jordan metric,
not the Pauli metric which describes the spin-2 graviton.
To obtain the metric which describes
the spin-2 graviton one need to make a proper conformal
transformation of the Jordan metric \cite{prl92}.

Notice that, when $\phi_{ij}$ becomes the bi-invariant
metric $\delta_{ij}$
\bea
&\phi_{ij}=\delta_{ij}=-\dfrac12 f_{ik}^{~~l} f_{jl}^{~~k},
\eea
the scalar source $\phi_{ij}$ disappears completely 
from (\ref{cact}), and the (4+n)-dimensional Einstein-Hilbert
action is simplified to
\bea
&S_P=-\dfrac{1}{16\pi G_N} \int {\sqrt -g} \Big(
R_M-\dfrac 14 \F_\mn^2 \nn\\
&+R_G-\Lambda_P \Big)~d^4 x,
\eea
where $R_M$ and $R_G$ are the scalar curvature of the
4-dimensional space-time and the n-dimensional internal space.
And $R_G$ (with $\phi_{ij}=\delta_{ij}$) becomes a constant.
This provides the (4+n)-dimensional
unification of gravitation and gauge theory with no other
source, where $R_G-\Lambda_P=\Lambda_M$ now plays the role of 
the $4$-dimensional cosmological constant \cite{jmp75,prd75}.

We emphasize that in this unification the volume of the
fiber space is space-time dependent, because $\phi=\phi(x)$.
More importantly, the volume of the group space $V_G$ 
(relative to the volume of the space-time $V_M$)
is left arbitrary, so that
the internal space can have any size. In spite of this
the internal space becomes completely unphysical
and unobservable. This is because
this unification is the unification by
isometry, in which the internal space becomes unobservable
because of the exact isometry \cite{jmp75,prd75,plb87}.
And there is no ``higher modes" of internal metric in this unification.
In comparison, in the popular higher-dimensional unification
where the internal space has no isometry,
the internal space becomes ``unobservable"
only when it is small (of the order of the Planck
scale) \cite{duff}. Moreover, one has to deal with
the ``higher modes" of internal metric in the popular unification.
We believe that our unification by isometry provides
the only logically consistent higher-dimensional
unification of gauge theory with gravitation \cite{pen}.

With this preliminary, we now discuss the geometry of
Abelian decomposition. Clearly any extra isometry
on $P$ will further restrict the metric $g_{AB}$,
and thus reduce the physical degrees of the unified theory.
Suppose we have an extra right isometry (the magnetic isometry)
$H$ which is vertical, which commutes with
the left isometry $G$. Let the right isometry be described
by $m_a$,
\bea
&\forall_a~~~{\cal L}_{m_a} g_{AB}=0, \nn\\
&m_a= m_a^{~i} \xi_i, \nn\\
&[m_a,~m_b]={\bar f}_{ab}^{~~c} m_c,  \nn\\
&[m_a,~\xi_i]=0.~~~(a,b,c=1,2,...,m<n)
\eea
Clearly the last commutation relation tells that
$\hm_a=(m_a^{~1},m_a^{~2},...,m_a^{~n})$ forms an adjoint
representation of $G$. Furthermore, when $\phi_{ij}=\delta_{ij}$,
the magnetic isometry can be written as \cite{prd80}
\bea
&\forall_a~~~D_\mu \hm_a =0,  \nn\\
&\hm_a \times \hm_b= \hm_c.
\label{miso}
\eea
This condition restricts the gauge potential, and defines the
restricted potential which satisfies the extra magnetic
isometry.

Let $H$ be the maximal Abelian subgroup of $G$, and let $\hA_\mu$ be
the restricted potential which satisfies the above condition. Then
the most general gauge potential $\A_\mu$ can be written as
\bea
&\A_\mu= \hA_\mu + \X_\mu,
\label{cdec}
\eea
where $\X_\mu$ is the $G/H$ part of
potential which transforms covariantly
under the gauge transformation.
This is because the space of gauge potentials (the
connection space) forms an Affine space, so that one can obtain
arbitrary potential by adding a gauge covariant vector field to the
restricted potential. This allows us to identify $\X_\mu$
as the valence potential. This decomposition of gauge potential is
called the Abelian decomposition \cite{prd80,prd02}.

The above Abelian decomposition has important features. First the
restricted potential $\hA_\mu$ enjoys
the full gauge symmetry of $G$, in spite
of the fact that it is restricted. This is because the left isometry
$G$ (the full gauge symmetry) remains intact in this
decomposition \cite{jmp75,prd75,plb87}.
Secondly, it retains all essential topological
properties of $G$. In particular, $\hA_\mu$ naturally
selects the magnetic potential which describes the non-Abelian monopole.
For this reason the right isometry (\ref{miso}) is called the magnetic
isometry. But most importantly, the decomposition is gauge
independent. Once $H$ is chosen, the decomposition follows
automatically, independent of the choice of a gauge.

\section{Abelian decomposition of $SU(2)$: An Example}

To show how the Abelian decomposition works, we consider the $SU(2)$
gauge theory for simplicity. Let $\hn$ be an arbitrary 
isotriplet unit vector field, and identify the maximal 
Abelian subgroup to be the $U(1)$ subgroup which leaves
$\hn$ invariant. Clearly $\hn$ selects
the ``Abelian'' direction (i.e., the color charge direction)
at each space-time point, and the the Abelian magnetic isometry
can be described by the following constraint equation
\bea
D_\mu \hn = \pro_\mu \hn + g \A_\mu \times \hn = 0.
~~~(\hn^2=1)
\eea
This has the unique solution for $\A_\mu$ which defines 
the restricted potential $\hA_\mu$
\bea
&\hA_\mu=A_\mu \hn - \oneg \hn\times\pro_\mu\hn,
\eea
where $A_\mu = \hn\cdot \A_\mu$ is the ``electric'' potential.
Notice that the restricted potential is precisely the connection
which leaves $\hn$ invariant under the parallel transport,
\bea
\hD_\mu \hn = \pro_\mu \hn + g \hA_\mu \times \hn = 0. \nn
\eea
This process of selecting the restricted potential is called the
Abelian projection \cite{prd80,prl81}.

With the Abelian projection we can retrieve the full gauge
potential by adding the gauge covariant valence potential $\X_\mu$
to the restricted potential,
\bea
& \A_\mu =A_\mu \hn - \oneg \hn\times\pro_\mu\hn+\X_\mu
= \hA_\mu + \X_\mu, \nn\\
&(\hn^2 =1,~~~ \hn\cdot\X_\mu=0).
\label{su2dec}
\eea
This is the Abelian decomposition which decomposes
the gauge potential into the restricted potential
$\hA_\mu$ and the valence potential $\X_\mu$ \cite{prd80,prl81}.

Let $\vec \alpha$ is an infinitesimal gauge parameter.
Under the infinitesimal gauge transformation
\bea
&\delta \hn = - \vec \alpha \times \hn,~~~~~~ \delta \vec A_\mu =
\oneg D_\mu \vec \alpha,
\label{su2gt}
\eea
one has
\bea
&\delta A_\mu = \oneg \hn \cdot \pro_\mu \vec \alpha,~~~~ \delta
\hA_\mu = \oneg \hD_\mu \vec \alpha, \nn\\
&\delta \X_\mu = - \vec \alpha \times \X_\mu.
\eea
This shows that $\hA_\mu$ by itself describes an $SU(2)$ connection
which enjoys the full $SU(2)$ gauge degrees of freedom. Furthermore
$\X_\mu$ transforms covariantly under the gauge transformation. Most
importantly, the decomposition is gauge-independent. Once the color
direction $\hn$ is selected, the decomposition follows independent
of the choice of a gauge. 

To understand the physical meaning of our decomposition notice that
the restricted potential $\hA_\mu$ actually has a dual structure.
Indeed the field strength made of the restricted potential is
decomposed as
\bea
&\hF_\mn=\pro_\mu \hA_\nu-\pro_\nu \hA_\mu+ g
\hA_\mu \times \hA_\nu  \nn\\
&=(F_\mn+ H_\mn)\hn, \nn\\
&F_\mn=\pro_\mu A_\nu-\pro_\nu A_\mu, \nn\\
&H_\mn=-\oneg \hn \cdot(\pro_\mu\hn\times\pro_\nu\hn) =\pro_\mu
\tC_\nu-\pro_\nu \tC_\mu,
\eea
where $\tC_\mu$ is the ``magnetic'' potential \cite{prd80,prl81}.
Notice that we can always introduce the magnetic potential (at least
locally section-wise), because $H_\mn$ forms a closed two-form
\bea
\pro_\mu H_\mn^d = 0 ~~~~~~~ ( H_\mn^d = \dfrac12
\epsilon_{\mn\rho\sigma} H_{\rho\sigma} ).
\eea
This allows us to identify the non-Abelian magnetic potential by
\bea
\C_\mu= -\oneg \hn \times \pro_\mu\hn ,
\label{mp}
\eea
in terms of which the magnetic field strength is expressed as
\bea
&\vec H_\mn=\pro_\mu \C_\nu-\pro_\nu \C_\mu+ g
\C_\mu \times \C_\nu  \nn\\
&=-g \C_\mu \times \C_\nu = -\oneg \pro_\mu\hn\times\pro_\nu\hn
=H_\mn\hn.
\eea
As importantly $\hA_\mu$, as an $SU(2)$ potential, retains all the
essential topological characteristics of the original non-Abelian
potential. This is because the topological field $\hn$ naturally
represents the non-Abelian topology $\pi_2(S^2)$ which describes the
mapping from an $S^2$ in 3-dimensional space $R^3$ to the coset space
$SU(2)/U(1)$, and $\pi_3(S^3)\simeq\pi_3(S^2)$ which describes the
mapping from the compactified 3-dimensional space 
$S^3$ to the group space $S^3$.
Clearly the isolated singularities of $\hn$ defines $\pi_2(S^2)$
which describes the non-Abelian monopoles.  Indeed $\C_\mu$ with
$\hn=\hat r$ describes precisely the Wu-Yang 
monopole \cite{prd80,prl80}. This is why we call $\C_\mu$ 
the magnetic potential.
Besides, with the $S^3$ compactification of $R^3$, $\hn$
characterizes the Hopf invariant $\pi_3(S^2)\simeq\pi_3(S^3)$ which
describes the topologically distinct vacua \cite{bpst,baal,plb06}.

With (\ref{su2dec}) we have
\bea
\F_\mn=\hF_\mn + \hD_\mu \X_\nu - \hD_\nu \X_\mu + g\X_\mu \times
\X_\nu,
\eea
so that the Yang-Mills Lagrangian is expressed as
\bea
&{\cal L}=-\dfrac{1}{4} \F^2_\mn =-\dfrac{1}{4} \hF_\mn^2
-\dfrac{1}{4} ( \hD_\mu \X_\nu -\hD_\nu \X_\mu)^2 \nn\\
&-\dfrac{g}{2} \hF_\mn \cdot (\X_\mu \times \X_\nu)
- \dfrac{g^2}{4} (\X_\mu \times \X_\nu)^2 \nn\\
&+ \lambda(\hn^2 -1) + \lambda_\mu \hn \cdot \X_\mu,
\label{su2lag}
\eea
where $\lambda$ and $\lambda_\mu$ are the Lagrangian multipliers.
From the Lagrangian we have
\bea
\delta A_\nu:&&\pro_\mu (F_\mn+H_\mn+X_\mn) \nn\\
&&= -g \hn \cdot \{ \X_\mu \times (\hD_\mu \X_\nu - \hD_\nu \X_\mu) \}, \nn\\
\delta \X_\nu:&&\hD_\mu ( \hD_\mu \X_\nu - \hD_\nu \X_\mu ) \nn\\
&&=g (F_\mn+H_\mn+X_\mn) \hn \times \X_\mu.\label{su2eq}
\eea
where
\bea
X_\mn = g \hn \cdot ( \X_\mu \times \X_\nu ). \nn
\eea
Notice that here $\hn$ has no equation of motion even though the
Lagrangian contains it explicitly. This is because it represents
a topological degrees of freedom, not a
local degrees of freedom \cite{prd80,prl81}.
From this we conclude that the non-Abelian gauge
theory can be viewed as a restricted gauge theory made of the
restricted potential, which has an additional colored source made of
the valence gluon.

Obviously the Lagrangian (\ref{su2lag}) is invariant under the
active gauge transformation (\ref{su2gt}). But notice that the
decomposition introduces another gauge symmetry that we call the
passive gauge transformation \cite{prd02,jhep04},
\bea
\delta \hn = 0, ~~~~~~~\delta \A_\mu = \oneg D_\mu \vec
\alpha,\label{su2gt2}
\eea
under which we have
\bea
&\delta A_\mu = \oneg \hn \cdot D_\mu \vec \alpha, ~~~~~~~\delta
\hA_\mu = \oneg (\hn \cdot D_\mu \vec \alpha) \hn, \nn\\
&\delta \X_\mu = \oneg \{ D_\mu \vec \alpha -(\hn \cdot D_\mu \vec
\alpha) \hn \}.
\eea
This is because, for a given $\A_\mu$, one can have infinitely many
different decomposition of (\ref{su2dec}), with different $\hA_\mu$
and $\X_\mu$ choosing different $\hn$. Equivalently, for a fixed
$\hn$, one can have infinitely many different $\A_\mu$ which are
gauge-equivalent to each other. So our decomposition automatically
induce another type of gauge invariance which comes from different
choices of decomposition. This extra gauge invariance plays a
crucial role in quantizing the theory \cite{prl81}.

An important advantage of the decomposition (\ref{su2dec})
is that it can actually ``Abelianize'' (or more precisely ``dualize'')
the non-Abelian dynamics, without any gauge fixing \cite{prd80,prd02}.
To see this let $(\hn_1,\hn_2,\hn_3=\hn)$ be 
a right-handed orthonormal basis and let
\bea
&\X_\mu =X^1_\mu ~\hn_1 + X^2_\mu ~\hn_2, \nn\\
&(X^1_\mu = \hn_1 \cdot \X_\mu,~~~X^2_\mu = \hn_2 \cdot \X_\mu) \nn
\eea
and find
\bea
&\hD_\mu \X_\nu =\{\pro_\mu X^1_\nu-g(A_\mu+ \tC_\mu)X^2_\nu \} \hn_1 \nn\\
&+ \{\pro_\mu X^2_\nu+ g (A_\mu+ \tC_\mu)X^1_\nu \}\hn_2,
\eea
where now the magnetic potential $\tC_\mu$
can be written explicitly as
\bea
&\tC_\mu=-\dfrac1g \n_1 \cdot \pro_\mu \n_2,
\eea
up to the $U(1)$ gauge transformation which leaves $\hn$ invariant.
So with
\bea
& \bA_\mu= A_\mu + \tC_\mu,
~~~\bF_\mn = \pro_\mu \bA_\nu - \pro_\nu \bA_\mu, \nn\\
&X_\mu = \dfrac{1}{\sqrt{2}} ( X^1_\mu + i X^2_\mu ),
\eea
one could express the Lagrangian explicitly in terms of the dual
potential $B_\mu$ and the complex vector field $X_\mu$,
\bea
&{\cal L}=-\dfrac{1}{4} \bF_\mn^2 -\dfrac12 |\bD_\mu X_\nu-\bD_\nu
X_\mu|^2 + ig \bF_\mn X_\mu^* X_\nu \nn\\
&-\dfrac12 g^2 \{(X_\mu^*X_\mu)^2-(X_\mu^*)^2 (X_\nu)^2\},
\label{su2lag2}
\eea
where now
\bea
\bD_\mu = \pro_\mu + ig \bA_\mu.  \nn
\eea
Clearly this describes an Abelian gauge theory coupled to the
charged vector field $X_\mu$. But the important point here is that
the Abelian potential $\bA_\mu$ is given by the sum of the electric
and magnetic potentials $A_\mu+ \tC_\mu$. In this form the equations
of motion (\ref{su2eq}) is re-expressed as
\bea
&\pro_\mu(\bF_\mn+X_\mn)
= i g X^*_\mu (\bD_\mu X_\nu - \bD_\nu X_\mu )  \nn\\
&- i g X_\mu (\bD_\mu X_\nu - \bD_\nu X_\mu )^* , \nn\\
&\bD_\mu(\bD_\mu X_\nu- \bD_\nu X_\mu)
=ig X_\mu (\bF_\mn+X_\mn).
\eea
where now
\bea
X_\mn = - i g ( X_\mu^* X_\nu - X_\nu^* X_\mu ) .  \nn
\eea
This shows that one can indeed Abelianize the non-Abelian theory
with our decomposition. The remarkable change in this ``Abelian''
formulation is that here the topological field $\hn$ is replaced by
the magnetic potential $\tC_\mu$.

One might ask whether this ``Abelian'' theory retains the original
non-Abelian gauge symmetry, and if so, how the non-Abelian gauge
symmetry is realized in this ``Abelian'' theory. To answer this
notice that here we have never fixed the gauge to obtain this
Abelian formalism, so that the original non-Abelian gauge symmetry
must remain intact. To see this let
\bea
&\vec \alpha = \alpha_1~\hn_1 + \alpha_2~\hn_2 + \theta~\hn, \nn\\
&\alpha = \dfrac{1}{\sqrt 2} (\alpha_1 + i ~\alpha_2), \nn\\
&\C_\mu = - \oneg \hn \times \pro_\mu \hn = - C^1_\mu \hn_1 - C^2_\mu \hn_2, \nn\\
&C_\mu = \dfrac{1}{\sqrt 2} (C^1_\mu + i ~ C^2_\mu).
\eea
Then the Lagrangian (\ref{su2lag2}) is invariant not only under the
active gauge transformation (\ref{su2gt}) described by
\bea
&\delta A_\mu = \oneg \pro_\mu \theta - i (C_\mu^* \alpha - C_\mu
\alpha^*),~~~\delta \tC_\mu = - \delta A_\mu, \nn\\
&\delta X_\mu = 0,\label{su2gt3}
\eea
but also under the passive gauge transformation (\ref{su2gt2})
described by
\bea
&\delta A_\mu = \oneg \pro_\mu \theta -i (X_\mu^* \alpha - X_\mu \alpha^*), ~~~\delta \tC_\mu = 0, \nn\\
&\delta X_\mu = \oneg \bD_\mu \alpha - i \theta X_\mu.\label{su2gt4}
\eea
This tells that the ``Abelian'' theory not only retains the original
non-Abelian gauge symmetry, but actually has an enlarged (both the
active and passive) gauge symmetries. Again this is because this
Abelianization is not the ``naive'' Abelianization of the $SU(2)$
gauge theory which one obtains by fixing the gauge. Our
Abelianization is a gauge independent Abelianization which retains
the full non-Abelian gauge symmetry of the original theory.
Moreover, we emphasize that here the Abelian gauge group is actually
made of $U(1)_e \otimes U(1)_m$, so that the Abelian theory becomes
a dual gauge theory \cite{prd80,prd02}. This is evident from
(\ref{su2gt3}) and (\ref{su2gt4}).

The above $U(1)$ decomposition of $SU(2)$ was first introduced
long time ago in an attempt to demonstrate the monopole condensation
and color confinement in QCD \cite{prd80,prl81}, 
and the importance of the decomposition in establishing 
the Abelian dominance in non-Abelian dynamics has become 
appreciated by many authors \cite{fadd,kondo}.
Now we show that this decomposition plays
a crucial role for us to make the Abelian decomposition of
the Einstein's theory.

\section{Decomposition of Gravitational Connection}

We apply the above Abelian decomposition to Einstein's theory,
regarding Einstein's theory as a gauge theory of Lorentz group.
To do this we introduce a coordinate
basis
\bea
[\pro_\mu,~\pro_\nu] = 0, ~~~(\mu,\nu =t,x,y,z) \nn
\eea
and an orthonormal basis
\bea
&[\xi_a,~\xi_b] = f_{ab}^{~~c} \xi_c. ~~~(a,b= 0,1,2,3) \nn \\
&\xi_a=e_a^{~\mu} \pro_\mu,~~~~\pro_\mu=e_\mu^{~a} \xi_a,
\eea
where $e_\mu^{~a}$ and $e_a^{~\mu}$ are the tetrad and inverse tetrad.
Let $J_{ab}=-J_{ba}$ be the generators of Lorentz group,
\bea
&[J_{ab}, ~J_{cd}] = \eta_{ac} J_{bd}-\eta_{bc} J_{ad}
+\eta_{bd} J_{ac}-\eta_{ad} J_{bc} \nn\\
&=f_{ab,cd}^{~~~~~~mn}~J_{mn}, \nn\\
&f_{ab,cd}^{~~~~~~mn}=\eta_{ac} \delta_b^{~[m} \delta_d^{~n]}
-\eta_{bc} \delta_a^{~[m} \delta_d^{~n]} \nn\\
&+\eta_{bd} \delta_a^{~[m} \delta_c^{~n]} -\eta_{ad} \delta_b^{~[m}
\delta_c^{~n]},
\label{lgcr}
\eea
where $\eta_{ab}=diag~(-1,1,1,1)$ is the Minkowski metric. Clearly
$J_{ab}$ has the following 4-dimensional matrix representation
\bea
&(J_{ab})_{c}^{~d}=-\eta_{ac}\delta_{b}^{~d}
+\eta_{bc}\delta_{a}^{~d},
\eea
so that under the infinitesimal gauge transformation we have
\bea
&\delta~e_\mu^c=(\eta_{ad}\delta_{b}^{~c}
-\eta_{bd}\delta_{a}^{~c})~\alpha^{ab}~e_\mu^d,
\eea
where $\alpha^{ab}(=-\alpha^{ba})$ is an infinitesimal gauge parameter of
the Lorentz group. Instead of $(ab,cd,...)$ we can use
the index $(A,B,...)=(1,2,3,4,5,6)=(23,31,12,01,02,03)$, and write
\bea
[J_A, ~J_B] ~~=~~ f_{AB}^{~~~C} ~J_C~.  \nn
\eea
Moreover, with
\bea
&L_{1,2,3}=J_{1,2,3}=J_{23,31,12} \nn\\
&K_{1,2,3}=J_{4,5,6}=J_{01,02,03} \nn
\eea
the Lorentz algebra is written as
\bea
& [L_i, ~L_j] = \epsilon_{ijk} L_k, \nn \\
& [L_i, ~K_j] = \epsilon_{ijk} K_k, \nn \\
& [K_i, ~K_j] =-\epsilon_{ijk} L_k, ~~~(i,j,k= 1,2,3)
\eea
where $L_i$ and $K_i$ are the 3-dimensional rotation and
boost generators. Notice that the generators can be viewed
as the left-invariant basis vector fields
on the Lorentz group manifold which satisfy
the commutation relation.

As we have pointed out, we can regard Einstein's theory as a gauge
theory of Lorentz group. In this view the gravitational connection
$\Gamma_\mn^{~~\rho}$ (or more precisely the spin connection
$\omega_\mu^{~ab}$) corresponds to the gauge potential
$\Gamma_\mu^{~ab}$, and the curvature tensor
$R_\mn^{~~ab}$ corresponds to the gauge field strength 
$F_\mn^{~~ab}$ of Lorentz group.
And to obtain the desired decomposition we have to decompose the
gauge potential $\Gamma_\mu^{~ab}$ first. Now, to apply the above $SU(2)$
decomposition to Lorentz group, we have to keep in mind that there
are notable differences between $SU(2)$ and Lorentz group. First,
the Lorentz group is non-compact, so that the invariant metric is
indefinite. Secondly, the Lorentz group has the well-known invariant
tensor $\epsilon_{abcd}$ which allows the dual transformation.
Thirdly, the Lorentz group has rank two, so that it has two
commuting Abelian subgroups and two Casimir invariants. 
Finally, the Lorentz group has two different maximal 
Abelian subgroups $A_2$ and $B_2$ \cite{jmp79}. 
These differences make the
decomposition more complicated.

The invariant metric $\delta_{AB}$ of Lorentz group
is given by
\bea
&\delta_{AB}= -\dfrac{1}{4} f_{AC}^{~~D}f_{BD}^{~~C} \nn\\
&=diag~(1,1,1,-1,-1,-1).
\label{inme}
\eea
Let $p^{ab}~(p^{ab}=-p^{ba})$ (or $p^A$) be a gauge covariant sextet
vector which forms an adjoint representation of Lorentz group,
\bea
\delta~p^{cd}= - \dfrac12 f_{ab,mn}^{~~~~~~cd} \alpha^{ab}~p^{mn}.
\eea
Clearly it can be understood as an anti-symmetric tensor in
4-dimensional Minkowski space which can be expressed by two
3-dimensional vectors $\m$ and $\e$, which transform exactly like
the magnetic and electric components of an electromagnetic tensor
under the 4-dimensional Lorentz transformation. And we can denote
$p^{ab}$ by $\vp$,
\bea
&\vp = \dfrac12 p_{ab} \vI^{ab}=\left( \begin{array}{c} \m \\
\e \end{array} \right),
~~p^{ab}=\vp \cdot \vI^{ab}=\dfrac12 p^{mn} I_{mn}^{~~~ab}, \nn\\
&\vI^{ab}=\left( \begin{array}{c} {\hat a}^{ab} \\
{\hat b}^{ab} \end{array} \right), \nn\\
&{\hat a}_i^{~ab}=\epsilon_{0i}^{~~ab}, ~~~{\hat
b}_i^{~ab}=\big(\delta_0^{~a} \delta_i^{~b} -\delta_0^{~b}
\delta_i^{~a} \big), \nn\\
&I_{cd}^{~~~ab}=\big(\delta_c^{~a} \delta_d^{~b} -\delta_c^{~b}
\delta_d^{~a} \big) =-(J_{cd})^{ab}.
\label{iddef}
\eea
where $m_i=\epsilon_{ijk}p^{jk}/2~(i,j,k=1,2,3)$ is the magnetic
(or rotation) part and $e_i=p^{0i}$  is the electric (or boost) part of
$\vp$. From the invariant metric (\ref{inme}) we have
\bea
\vp^2= \dfrac 12 p_{ab} p^{ab}= \m^2-\e^2,
\eea
so that the invariant length can be positive, zero, or negative.
This, of course, is due to the fact that the invariant metric
(\ref{inme}) is indefinite.

The Lorentz group has another important invariant tensor
$\epsilon_{AB}$ which comes from the totally anti-symmetric
invariant tensor $\epsilon_{abcd}$,
\bea
\epsilon_{AB}=\epsilon_{ab,cd}=\epsilon_{abcd}.
\eea
This tells that any adjoint representation of Lorentz group has its
dual partner. In particular, $\vp$ has the dual vector $\vtp$
defined by $\widetilde p^{ab}=\epsilon^{abcd} p_{cd}/2$. With
(\ref{iddef}) we have (with $\epsilon_{0123}=+1$)
\bea
&\vtp=\left( \begin{array}{c} \e \\ -\m
\end{array} \right),
~~~~~\tilde{\vtp}=-\vp, \nn\\
&\vtp^2=\e^2-\m^2=-\vp^2,  \nn\\
&\vp \cdot \vtp= \dfrac{1}{4} \epsilon_{abcd} p^{ab} p^{cd}
=2 \m \cdot \e. 
\eea
Moreover, since
\bea
&(p \times p')^{ab}=(\vp \times \vp') \cdot \vI^{ab} \nn\\
&=- \eta_{cd}(p^{ac} p'^{bd}-p^{bc} p'^{ad}),
\eea
we have
\bea
[~p, ~\widetilde p~]=0,
\label{cl}
\eea
or equivalently
\bea
{\vp \times \vtp} =0
\eea
This tells that any two vectors which are dual to each other are
always commuting. Finally with two vectors $\vp$ and $\vp'$ we have
\bea
&{\vp \cdot \vp'}= \m \cdot \m'-\e \cdot \e', \nn\\
&{\vp \cdot \vtp'}= \m \cdot \e'+\e \cdot \m'={\vtp \cdot \vp'}, \nn\\
&\vp \times \vp'=\left(\begin{array}{c} \m \times \m'-\e \times \e' \\
\m \times \e'+\e \times \m'  \end{array} \right)
=- \vtp \times \vtp', \nn\\
&\vp \times \vtp'= \left(\begin{array}{c} \m \times \e' +\e
\times\m' \\ -\m \times \m' +\e \times \e'
\end{array} \right) = \vtp \times \vp', \nn\\
&\widetilde{\vp \times \vp'}=\vp \times \vtp',
\eea
so that we can always reduce the operations of sextet vectors
of Lorentz group to the operations of 3-dimensional vectors.

Let $(\hn_1,\hn_2,\hn_3=\hn)$ be a $3$-dimensional unit vectors
($\hn_i^2=1$) which form a right-handed orthonormal
basis with $\hn_1 \times \hn_2=\hn_3$, and let
\bea
&\vl_i= \left( \begin{array}{c} \hn_i \\
0 \end{array} \right),
~~~\vk_i= \left( \begin{array}{c} 0 \\
\hn_i  \end{array} \right)= -\vtl_i. \label{lbasis}
\eea
Clearly we have
\bea
&\vl_i \cdot \vl_j=\delta_{ij}, ~~~\vl_i \cdot \vk_j=0,
~~~\vk_i\cdot \vk_j=-\delta_{ij},  \nn\\
&\vl_i \times \vl_j= \epsilon_{ijk} \vl_k, \nn\\
&\vl_i \times \vk_j= \epsilon_{ijk} \vk_k, \nn\\
&\vk_i \times \vk_j= -\epsilon_{ijk} \vl_k
\eea
so that $(\vl_i,\vk_i)$, or equivalently $(\vl_i,\vtl_i)$,
forms an orthonormal basis of
the adjoint representation of Lorentz group.

To make the desired Abelian decomposition we have to choose the gauge
covariant sextet vector fields which form adjoint representation of
Lorentz group which describe the desired magnetic isometry. To see
what types of isometry is possible, it is important to
remember that Lorentz group has
two maximal Abelian subgroups, $A_2$ made of $L_3$ and $K_3$ and
$B_2$ made of $(L_1+K_2)/\sqrt 2$ and $(L_2-K_1)/\sqrt 2$
\cite{jmp79}. This tells that we have two possible 
Abelian decompositions of the gravitational connection. 
And in both cases the magnetic isometry is
described by two, not one, commuting sextet vector fields of Lorentz
group which are dual to each other. To see this let us denote one of
the isometry vector field by $\vp$ which satisfy the isometry
condition
\bea
D_\mu \vp = (\pro_\mu + \vGm_\mu \times) ~\vp=0, \label{ic}
\eea
where we have normalized the coupling constant to be the unit
(which one can always do without loss of generality).
Now, notice that the above condition automatically assures
\bea
D_\mu \vtp =(\pro_\mu + \vGm_\mu \times) ~\vtp=0, \label{dic}
\eea
because $\epsilon_{abcd}$ is an invariant tensor. This tells that
when $\vp$ is an isometry, $\vtp$ also becomes an isometry. To
verify this directly we decompose the gauge potential of Lorentz
group $\vGm_\mu$ into the 3-dimensional rotation and boost parts
$\A_\mu$ and $\B_\mu$, and let
\bea
\vGm_\mu= \left( \begin{array}{c} \A_\mu \\
\B_\mu \end{array} \right).
\eea
With this both (\ref{ic}) and (\ref{dic}) can be written as
\bea
&D_\mu \m= \B_\mu \times \e, \nn\\
&D_\mu \e= -\B_\mu \times \m,
\label{ic1}
\eea
where now
\bea
D_\mu = \pro_\mu + \A_\mu \times. \nn
\eea
This confirms that (\ref{ic}) and (\ref{dic}) are actually identical
to each other, which tells that the magnetic isometry in Lorentz
group must be even-dimensional.

Since Lorentz group has two invariant tensors it has two Casimir
invariants. And it is useful to characterize the isometry by two
Casimir invariants. Let the isometry is described by $\vp$ and
$\vtp$. It has two Casimir invariants $\alpha$ and $\beta$,
\bea
&\alpha ~=~ {\vp \cdot \vp} ~=~ \m^2-\e^2, \nn\\
&\beta ~=~ {\vp \cdot \vtp} ~=~2 \m \cdot \e.
\eea
But the Casimir invariants $(\alpha,\beta)$
depends on the choice of the isometry vectors.
To see this consider $\vp'$ and $\vtp'$ given by
a linear combination of $\vp$ and $\vtp$,
\bea
&\vp'= a\vp+b\vtp,
~~~~~\vtp'=a\vtp -b\vp.
\eea
Clearly we have
\bea
&D_\mu \vp'=0,~~~~~D_\mu \vtp'=0,
\eea
so that they can also be viewed to describe the same isometry.
But their Casimir invariants $(\alpha',\beta')$
are given by
\bea
&\alpha'= (a^2-b^2)\alpha+2ab\beta, \nn\\
&\beta'=(a^2-b^2)\beta-2ab\alpha.
\eea
And with
\bea
&a^2=\dfrac{\sqrt{\alpha^2+\beta^2}+|\alpha|}{2(\alpha^2+\beta^2)},
~~~b^2=\dfrac{\sqrt{\alpha^2+\beta^2}-|\alpha|}{2(\alpha^2+\beta^2)}, \nn
\eea
we can always make
\bea
&\alpha'= 1,~~~~~\beta'= 0,
\eea
unless $\alpha^2+\beta^2=0$. This tells that
we can always choose $\vp$ and $\vtp$ in such a way
to make $(\alpha,\beta)$ to be $(1,0)$
or $(0,0)$. Physically this means that the magnetic isometry
in Einstein's theory can be classified by
the space-like (or equivalently time-like) isometry
and the light-like isometry whose Casimir invariants 
are denoted by $(1,0)$ and $(0,0)$, respectively.
Notice that there is no need to talk about the time-like isometry,
because the isometry vectors always contains the dual partners. 
But we emphasize that once $\vp$ and $\vtp$ are chosen,
$(\alpha,\beta)$ are uniquely fixed.
Now we discuss the two isometries
$A_2$ and $B_2$ separately.

\subsection{$A_2$ (Space-like) Isometry}

Let the maximal Abelian subgroup be $A_2$.
In this case the isometry is made of $L_3$ and $K_3$, and we have
two sextet vector fields which describes the isometry which are dual
to each other. Let $\vp$ and $\vtp$ be the two isometry vector
fields which correspond to $L_3$ and $K_3$. Clearly we can put
\bea
&\vp= f~\vl_3= f \left( \begin{array}{c} \hn \\
0  \end{array} \right),  \nn\\
&\vtp=f~\vtl_3= f \left( \begin{array}{c} 0 \\
-\hn  \end{array} \right),
\eea
where $f$ is an arbitrary function of space-time.
The Casimir invariants of the isometry vectors 
are given by $(f^2,0)$.
But just as in $SU(2)$ gauge theory the isometry
condition (\ref{ic}) requires $f$ to be a constant,
because
\bea
\pro_\mu f^2=\pro_\mu \vp^2=2\vp \cdot D_\mu \vp=0.
\eea
And we can always normalize $f=1$ without loss of generality.

So the $A_2$ isometry can always be written as
\bea
&\vl=\vl_3= \left( \begin{array}{c} \hn \\
0  \end{array} \right),
~~~~~\vtl=\vtl_3=\left( \begin{array}{c} 0 \\ -\hn
\end{array} \right), \nn\\
&{D_\mu \vl} =0,~~~~~{D_\mu \vtl}=0,
\label{a2ic}
\eea
whose Casimir invariants are fixed by $(1,0)$.
With this we find the restricted connection $\hvGm_\mu$
which satisfies the isometry condition
\bea
&\hvGm_\mu= \Gamma_\mu ~\vl - \tGm_\mu ~\vtl
-\vl \times \pro_\mu \vl, \nn\\
&\Gamma_\mu = {\vl \cdot \vGm_\mu},
~~~\tGm_\mu  = \vtl \cdot \vGm_\mu,
\label{a2rc}
\eea
where $\Gamma_\mu$ and $\tGm_\mu$ are two Abelian connections
of $\vl$ and $\vtl$ components which are not restricted by
the isometry condition. At first glance this expression
appears strange, because one expects that $\vl$ and $\vtl$ should
contribute equally in the restricted connection since (\ref{ic}) and
(\ref{dic}) are identical. Actually they do contribute equally
because we have
\bea
&\vl \times \pro_\mu \vl=-\vtl \times \pro_\mu \vtl,
\eea
so that we can express the restricted connection as
\bea
&\hvGm_\mu= \Gamma_\mu ~\vl - \tGm_\mu ~\vtl
-\dfrac12(\vl \times \pro_\mu \vl-\vtl \times \pro_\mu \vtl).
\eea
The restricted field strength $\hvR_\mn$ which describes
the restricted curvature tensor $\hR_\mn^{~~ab}$ is given by
\bea
&\hvR_\mn=\pro_\mu \hvGm_\nu-\pro_\nu \hvGm_\mu
+\hvGm_\mu \times \hvGm_\nu \nn\\
&= (\Gamma_\mn+ H_\mn) ~\vl - (\tGm_\mn+\tH_\mn) ~\vtl, \nn\\
&\Gamma_\mn = \pro_\mu \Gamma_\nu - \pro_\nu \Gamma_\mu,
~~~\tGm_\mn= \pro_\mu \tGm_\nu - \pro_\nu \tGm_\mu, \nn\\
&H_\mn = -\vl \cdot (\pro_\mu \vl \times \pro_\nu \vl), \nn\\
&\tH_\mn =-\vtl \cdot (\pro_\mu \vl \times \pro_\nu \vl)
=\vtl \cdot (\pro_\mu \vtl \times \pro_\nu \vtl)  \nn\\
&=0,
\eea
so that we have
\bea
&\hR_\mn^{~~ab}=\hvR_\mn \cdot \vI^{ab} \nn\\
&= (\Gamma_\mn+ H_\mn) ~l^{ab} - \tGm_\mn ~\tl^{ab}.
\label{a2rct}
\eea
Notice that $\tH_\mn$ vanishes.

In 3-dimensional notation the isometry condition (\ref{a2ic}) can be
written as
\bea
&\hat \vGm_\mu=\left( \begin{array}{c} \hat A_\mu \\
\hat B_\mu  \end{array} \right), \nn\\
&\hD_\mu \hn=0,~~~~~\hB_\mu \times \hn=0, \nn\\
&\hD_\mu=\pro_\mu+\hA_\mu \times.
\eea
From this we have
\bea
&\hat A_\mu=\Gamma_\mu \hn - \hn \times \pro_\mu \hn,
~~~\hat B_\mu=\tGm_\mu \hn, \nn\\
&\Gamma_\mu=\hn \cdot \hat A_\mu,~~~\tGm_\mu=\hn \cdot \hat B_\mu.
\label{a2rc3}
\eea
Moreover, with
\bea
\hvR_\mn=\left( \begin{array}{c} \hA_\mn \\
\hB_\mn  \end{array} \right), \label{rct}
\eea
we have
\bea
&\hA_\mn=\pro_\mu \hA_\nu-\pro_\nu \hA_\mu
+\hA_\mu \times \hA_\nu-\hB_\mu \times \hB_\nu \nn\\
&=(\Gamma_\mn + H_\mn) \hn, \nn\\
&\hB_\mn=\pro_\mu \hB_\nu-\pro_\nu \hB_\mu
+\hA_\mu \times \hB_\nu+\hB_\mu \times \hA_\nu \nn\\
&=\hD_\mu \hB_\nu-\hD_\nu \hB_\mu =\tGm_\mn ~\hn, \nn\\
&H_\mn=-\hn \cdot (\pro_\mu \hn \times \pro_\nu \hn)
=\pro_\mu \tC_\nu - \pro_\nu \tC_\mu, \nn\\
&\tC_\mu= -\hn_1 \cdot \pro_\mu \hn_2.
\eea
Notice that $\hA_\mu$ and $\hA_\mn$ are formally identical to the
restricted potential and restricted field strength of $SU(2)$ gauge
theory. In particular $H_\mn$ is identical to what we have in
Section III. This, together with $\tH_\mn =0$, tells that
the topology of this isometry is identical to that of the $SU(2)$
subgroup.

With this the full connection of Lorentz group is given by
\bea
&\vGm_\mu = \hvGm_\mu + \vZ_\mu, ~~~{\vl \cdot \vZ_\mu = \vtl \cdot
\vZ_\mu} = 0,
\eea
where $\vZ_\mu$ is the valence connection which transforms
covariantly under the Lorentz gauge transformation,
or equivalently under the general coordinate transformation.
The corresponding field strength $\vR_\mn$ which describes
the curvature tensor is written as
\bea
&\vR_\mn = \pro_\mu \vGm_\nu-\pro_\nu \vGm_\mu
+\vGm_\mu \times \vGm_\nu \nn\\
&=\hvR_\mn+\vZ_\mn, \nn\\
&\vZ_\mn=\hD_\mu \vZ_\nu - \hD_\nu \vZ_\mu
+ \vZ_\mu \times \vZ_\nu, \nn\\
&\hD_\mu=\pro_\mu + \hvGm_\mu \times, \label{dct}
\eea
where $\vZ_\mn$ is the valence part of the curvature tensor
which can further be decomposed to the kinetic part $\dZ_\mn$
and the potential part $\pZ_\mn$,
\bea
&\vZ_\mn=\dZ_\mn+\pZ_\mn, \nn\\
&\dZ_\mn=\hD_\mu \vZ_\nu - \hD_\nu \vZ_\mu,
~~~\pZ_\mn=\vZ_\mu \times \vZ_\nu.
\eea
Now with
\bea
&\vZ_\mu = \Za_\mu \vl_1 -\Zb_\mu \vtl_1
+\Zc_\mu \vl_2 -\Zd_\mu \vtl_2, \nn\\
&\Za_\mu =\vl_1 \cdot \vZ_\mu,~~~\Zb_\mu =\vtl_1 \cdot \vZ_\mu, \nn\\
&\Zc_\mu =\vl_2 \cdot \vZ_\mu,~~~\Zd_\mu =\vtl_2 \cdot \vZ_\mu,
\eea
we have
\bea
&\dZ_\mn= (\cD_\mu \Za_\nu-\cD_\nu \Za_\mu)\vl_1
-(\cD_\mu \Zb_\nu-\cD_\nu \Zb_\mu)\vtl_1 \nn\\
&+(\cD_\mu \Zc_\nu-\cD_\nu \Zc_\mu)\vl_2
-(\cD_\mu \Zd_\nu-\cD_\nu \Zd_\mu)\vtl_2, \nn\\
&\cD_\mu \Za_\nu = \pro_\mu \Za_\nu -L_\mu \Zc_\nu
-\tL_\mu \Zd_\nu,\nn\\
&\cD_\mu \Zb_\nu = \pro_\mu \Zb_\nu -L_\mu \Zd_\nu
+\tL_\mu \Zc_\nu,\nn\\
&\cD_\mu \Zc_\nu = \pro_\mu \Zc_\nu +L_\mu \Za_\nu
-\tL_\mu \Zb_\nu, \nn\\
&\cD_\mu \Zd_\nu = \pro_\mu \Zd_\nu +L_\mu \Zb_\nu
+\tL_\mu \Za_\nu, \nn\\
&L_\mu =\Gamma_\mu+\tC_\mu,~~~\tL_\mu=\tGm_\mu, \nn\\
&\vl \cdot \dZ_\mn=\vtl \cdot \dZ_\mn=0.
\eea
Clearly $L_\mu$ is identical to the dual potential
we have introduced in Section III in $SU(2)$ gauge theory.
Moreover, we have
\bea
&\pZ_\mn=W_\mn \vl - \tW_\mn \vtl, \nn\\
& W_\mn = {\vl \cdot (\vZ_\mu \times \vZ_\nu)} \nn\\
&=\Za_\mu\Zc_\nu-\Za_\nu\Zc_\mu -\Zb_\mu\Zd_\nu+\Zb_\nu\Zd_\mu, \nn\\
&\tW_\mn = \vtl \cdot (\vZ_\mu \times \vZ_\nu)  \nn\\
&=\Za_\mu\Zd_\nu-\Za_\nu\Zd_\mu +\Zb_\mu\Zc_\nu-\Zb_\nu\Zc_\mu.
\eea
With this we have the full curvature tensor
\bea
&\vR_\mn = (\Gamma_\mn+ H_\mn+ W_\mn)~{\vl}
-(\tGm_\mn+ \tW_\mn)~{\vl} \nn\\
&+ \hD_\mu \vZ_\nu - \hD_\nu \vZ_\mn \nn\\
&=(\cD_\mu L_\nu-\cD_\nu L_\mu)~{\vl}
-(\cD_\mu \tL_\nu-\cD_\nu \tL_\mu)~\vtl \nn\\
&+(\cD_\mu \Za_\nu-\cD_\nu \Za_\mu)~\vl_1
-(\cD_\mu \Zb_\nu-\cD_\nu \Zb_\mu)~\vtl_1 \nn\\
&+(\cD_\mu \Zc_\nu-\cD_\nu \Zc_\mu)~\vl_2
-(\cD_\mu \Zd_\nu-\cD_\nu \Zd_\mu)~\vtl_2 \nn\\
&=L_\mn^1~\vl_1 -\tL_\mn^1~\vtl_1
+L_\mn^2~\vl_2 -\tL_\mn^2~\vtl_2 \nn\\
&+ L_\mn ~{\vl} - \tL_\mn ~\vtl,  \nn\\
&\cD_\mu L_\nu=\pro_\mu L_\nu+\Za_\mu\Zc_\nu-\Zb_\mu\Zd_\nu,  \nn\\
&\cD_\mu \tL_\nu=\pro_\mu \tL_\nu-\Za_\mu\Zd_\nu-\Zb_\mu\Zc_\nu, \nn\\
&L_\mn^1=\cD_\mu \Za_\nu-\cD_\nu \Za_\mu,
~~~\tL_\mn^1=\cD_\mu \Zb_\nu-\cD_\nu \Zb_\mu, \nn\\
&L_\mn^2=\cD_\mu \Zc_\nu-\cD_\nu \Zc_\mu,
~~~\tL_\mn^2=\cD_\mu \Zd_\nu-\cD_\nu \Zd_\mu,  \nn\\
&L_\mn=\cD_\mu L_\nu-\cD_\nu L_\mu
=\Gamma_\mn+ H_\mn+ W_\mn, \nn\\
&\tL_\mn=\cD_\mu \tL_\nu-\cD_\nu \tL_\mu=\tGm_\mn+ \tW_\mn,
\label{a2ct1}
\eea
or equivalently
\bea
&R_\mn^{~~ab}=\vR_\mn \cdot \vI^{ab}  \nn\\
&=L_\mn^1~l_1^{ab}-\tL_\mn^1~\tl_1^{ab}
+L_\mn^2~l_2^{ab}-\tL_\mn^2~\tl_2^{ab}  \nn\\
&+L_\mn ~l^{ab} - \tL_\mn ~\tl^{ab}.
\label{a2ct2}
\eea
This is the $A_2$ decomposition of the curvature tensor.
The similarity between this decomposition and the
Abelian decomposition of $SU(2)$ is unmistakable.

To emphasize the similarity between this isometry and the
$U(1)$ isometry of $SU(2)$ we introduce the complex notation
\bea
&Z_\mu=\dfrac{1}{\sqrt2}(\Za_\mu+i\Zc_\mu),
~~~~ \tZ_\mu= \dfrac{1}{\sqrt2}(\Zb_\mu+i\Zd_\mu), \nn\\
&\vl_\pm=\dfrac{1}{\sqrt2}(\vl_1\pm i\vl_2),
~~~~\vtl_\pm=\dfrac{1}{\sqrt2}(\vtl_1\pm i\vtl_2),
\eea
and find
\bea
&\dZ_\mn= (\cD_\mu Z_\nu-\cD_\nu Z_\mu)^*~\vl_+
+(\cD_\mu Z_\nu-\cD_\nu Z_\mu)~\vl_-  \nn\\
&-(\cD_\mu \tZ_\nu-\cD_\nu \tZ_\mu)^*~\vtl_+
-(\cD_\mu \tZ_\nu-\cD_\nu \tZ_\mu)~\vtl_-, \nn\\
&\cD_\mu Z_\nu =(\pro_\mu +i L_\mu) Z_\nu
-i \tL_\mu \tZ^*_\nu=\bD_\mu Z_\nu-i \tL_\mu \tZ^*_\nu, \nn\\
&\bD_\mu = \pro_\mu +i L_\mu.
\eea
Here $\bD_\mu$ is identical to the one we have in Section III.
Moreover, the potential part of $\vZ_\mn$ is given by
\bea
&\pZ_\mn= W_\mn\vl-\tW_\mn\vtl, \nn\\
&W_\mn = \Za_\mu\Zc_\nu-\Za_\nu\Zc_\mu
-\Zb_\mu\Zd_\nu+\Zb_\nu\Zd_\mu \nn\\
&=-i(Z^*_\mu Z_\nu-Z^*_\nu Z_\mu -\tZ^*_\mu \tZ_\nu+\tZ^*_\nu \tZ_\mu), \nn\\
&\tW_\mn = \Za_\mu\Zd_\nu-\Za_\nu\Zd_\mu
+\Zb_\mu\Zc_\nu-\Zb_\nu\Zc_\mu \nn\\
&=-i(Z^*_\mu \tZ_\nu-Z^*_\nu \tZ_\mu -\tZ^*_\mu Z_\nu+\tZ^*_\nu
Z_\mu).
\eea
With this we have
\bea
&\vR_\mn = (\cD_\mu Z_\nu-\cD_\nu Z_\mu)^*~\vl_+
-(\cD_\mu \tZ_\nu-\cD_\nu \tZ_\mu)^*~\vtl_+  \nn\\
&+(\cD_\mu Z_\nu-\cD_\nu Z_\mu)~\vl_-  
-(\cD_\mu \tZ_\nu-\cD_\nu \tZ_\mu)~\vtl_-  \nn\\
&+(\cD_\mu L_\nu-\cD_\nu L_\mu)~\vl
-(\cD_\mu \tL_\nu-\cD_\nu \tL_\mu)~\vtl,
\label{a2ct3}
\eea
or
\bea
&R_\mn^{~~ab}=(\cD_\mu Z_\nu-\cD_\nu Z_\mu)^*~l_+^{ab}
-(\cD_\mu \tZ_\nu-\cD_\nu \tZ_\mu)^*~\tl_+^{ab} \nn\\
&+(\cD_\mu Z_\nu-\cD_\nu Z_\mu)~l_-^{ab}
-(\cD_\mu \tZ_\nu-\cD_\nu \tZ_\mu)~\tl_-^{ab}  \nn\\
&+(\cD_\mu L_\nu-\cD_\nu L_\mu)~l^{ab}
-(\cD_\mu \tL_\nu-\cD_\nu \tL_\mu)~\tl^{ab}.
\label{a2ct4}
\eea
This should be compared with the $SU(2)$ decomposition.

In 3-dimensional notation we have
\bea
&\vZ_\mu=\left( \begin{array}{c} \X_\mu \\
\Y_\mu  \end{array} \right), \nn\\
&\X_\mu=Z_\mu^1~\hn_1+ Z_\mu^2~\hn_2,
~~~\Y_\mu=\tZ_\mu^1~\hn_1+ \tZ_\mu^2~\hn_2,  \nn\\
&\hn \cdot \X_\mu=0,~~~~~\hn \cdot \Y_\mu=0.
\label{3dvc}
\eea
Moreover, with
\bea
&\vZ_\mn=\left( \begin{array}{c} \X_\mn \\
\Y_\mn  \end{array} \right) =\left( \begin{array}{c} \dX_\mn+\pX_\mn \\
\dY_\mn+\pY_\mn  \end{array} \right),
\eea
we have
\bea
&\dX_\mn=\hD_\mu \X_\nu -\hD_\nu \X_\mu
-\B_\mu \times \Y_\nu+\B_\nu \times \Y_\mu \nn\\
&= L_\mn^1~\hn_1+ L_\mn^2~\hn_2, \nn\\
&\dY_\mn=\hD_\mu \Y_\nu -\hD_\nu \Y_\mu
-\B_\mu \times \X_\nu -\B_\nu \times \X_\mu \nn\\
&=\tL_\mn^1~\hn_1+ \tL_\mn^2~\hn_2, \nn\\
&\pX_\mn= \X_\mu \times \X_\nu-\Y_\mu \times \Y_\nu =W_\mn~\hn, \nn\\
&\pY_\mn= \X_\mu \times \Y_\nu+\Y_\mu \times \X_\nu =\tW_\mn~\hn.
\eea
Notice that the kinetic part and the potential part of $\vZ_\mn$ are
orthogonal to each other.
Finally, with
\bea
&\vR_\mn=\left( \begin{array}{c} \A_\mn \\
\B_\mn  \end{array} \right)=\left( \begin{array}{c} \hA_\mn+\X_\mn \\
\hB_\mn+\Y_\mn  \end{array} \right),
\label{fct}
\eea
we have
\bea
&\A_\mn=L_\mn ~\hn +\dX_\mn \nn\\
&=L_\mn^1~\hn_1+L_\mn^2~\hn_2+L_\mn ~\hn, \nn\\
&\B_\mn=\tL_\mn ~\hn +\dY_\mn \nn\\
&=\tL_\mn^1~\hn_1+\tL_\mn^2~\hn_2+\tL_\mn ~\hn.
\eea
This completes the $A_2$ decomposition of the gravitational
connection.

\subsection{$B_2$ (Light-like) Isometry}

This is when the isometry group is made of $(L_1+K_2)/\sqrt 2$ and
$(L_2-K_1)/\sqrt 2$.
Let $\vp$ and $\vtp$ be the two isometry vector fields which
correspond to $(L_1+K_2)/\sqrt 2$ and $(L_2-K_1)/\sqrt 2$ which are
dual to each other. In this case we can write
\bea
&{\vp}= f\Big(\dfrac{\vl_1+ \vk_2}{\sqrt 2}\Big)
=\dfrac{f}{\sqrt 2} \left( \begin{array}{c} \hn_1 \\
 \hn_2  \end{array} \right), \nn\\
&{\vtp}=f\Big(\dfrac{ \vl_2-\vk_1}{\sqrt 2}\Big)
=\dfrac{f}{\sqrt 2} \left( \begin{array}{c}  \hn_2 \\
- \hn_1  \end{array} \right).
\eea
But notice that the Casimir invariants ($\alpha,\beta$) 
of the isometry vectors are given by
($0,0$) independent of $f$. Moreover, here
(unlike the $A_2$ case) the isometry
condition does not restrict $f$ at all, because we have
\bea
&\pro_\mu \vp^2=2 \vp \cdot D_\mu\vp=0,
\eea
independent of $f$. So the $B_2$ isometry vectors contain
an arbitrary scalar function $f(x)$.

Let us put $f=e^{\lambda}$ and express the $B_2$ isometry by
\bea
&{\vj}= \dfrac{e^\lambda}{\sqrt2} (\vl_1+ \vk_2)
=\dfrac{e^\lambda}{\sqrt2} \left(\begin{array}{c} \hn_1 \\
\hn_2  \end{array} \right),  \nn\\
&{\vtj}= \dfrac{e^\lambda}{\sqrt2} (\vl_2-\vk_1)
=\dfrac{e^\lambda}{\sqrt2} \left( \begin{array}{c} \hn_2 \\
-\hn_1  \end{array} \right), \nn\\
&D_\mu \vj=0,~~~D_\mu \vtj=0,
\label{b2ic}
\eea
To find the restricted connection $\hvGm$ which satisfies the
isometry condition we first introduce 4 more basis vectors in Lorentz
group manifold which together with $\vj$ and $\vtj$
form a complete basis
\bea
&\vk= \dfrac{e^{-\lambda}}{\sqrt 2} (\vl_1 -\vk_2)
=\dfrac{e^{-\lambda}}{\sqrt2} \left( \begin{array}{c} \hn_1 \\
-\hn_2 \end{array} \right),\nn\\
&\vtk= -\dfrac{e^{-\lambda}}{\sqrt 2} (\vl_2 +\vk_1)
=\dfrac{e^{-\lambda}}{\sqrt2} \left(\begin{array}{c}
-\hn_2 \\ -\hn_1 \end{array}\right), \nn\\
&\vl=-\vj \times \vtk=-\vtj \times \vk = \left( \begin{array}{c} \hn_3 \\
0 \end{array} \right),\nn\\
&\vtl=\vj \times \vk=-\vtj \times \vtk = \left( \begin{array}{c} 0 \\
-\hn_3 \end{array}\right).
\eea
Notice that 4 of them are null vectors,
\bea
&{\vj}^2= {\vtj}^2=\vk^2=\vtk^2=0,
\eea
but we have
\bea
&\vj \cdot \vk = -\vtj \cdot \vtk = 1,
~~~\vl^2=-\vtl^2=1.
\eea
All other scalar products of the basis vectors vanish.
Moreover we have
\bea
&\vj \times \vl= -\vtj \times \vtl =-\vtj,
~~~\vtj \times \vl=\vj \times \vtl=\vj, \nn\\
&\vk \times \vl= -\vtk \times \vtl =\vtk,
~~~\vtk \times \vl=\vk \times \vtl=-\vk.
\eea
From this we find the following restricted connection
for the $B_2$ isometry,
\bea
&\hvGm_\mu =\Gamma_\mu~\vj - \tGm_\mu~\vtj - \vk
\times \pro_\mu \vj  \nn\\
&= \Gamma_\mu~\vj - \tGm_\mu~\vtj
- \dfrac12(\vk \times \pro_\mu \vj-\vtk \times \pro_\mu, \vtj) \nn\\
&\Gamma_\mu = \vk \cdot \vGm_\mu,
~~~\tGm_\mu = \vtk \cdot \vGm_\mu, \nn\\
&\vk \times \pro_\mu \vj=-\vtk \times \pro_\mu \vtj~,
\label{b2rc}
\eea
where $\Gamma_\mu$ and $\tGm_\mu$ are two Abelian connections
of $\vj$ and $\vtj$ components which are not restricted
by the isometry condition.

The restricted curvature tensor $\hvR_\mn$ is given by
\bea
&\hvR_\mn=\pro_\mu \hvGm_\nu-\pro_\nu \hvGm_\mu
+\hvGm_\mu \times \hvGm_\nu \nn\\
&=(\Gamma_\mn+H_\mn) \vj
-(\tGm_\mn+\tH_\mn) \vtj, \nn\\
& \Gamma_\mn = \pro_\mu \Gamma_\nu - \pro_\nu \Gamma_\mu,
~~~\tGm_\mn = \pro_\mu \tGm_\nu- \pro_\nu \tGm_\mu, \nn\\
& H_\mn = -\vk \cdot (\pro_\mu \vj \times \pro_\nu \vk
-\pro_\nu \vj \times \pro_\mu \vk), \nn\\
& \tH_\mn= -\vtk \cdot (\pro_\mu \vj \times \pro_\nu \vk -\pro_\nu
\vj \times \pro_\mu \vk),
\eea
so that
\bea
&\hR_\mn^{~~ab}=(\Gamma_\mn+H_\mn) j^{ab} -(\tGm_\mn+\tH_\mn)
\tj^{ab}. 
\label{b2rct}
\eea
This should be contrasted with the restricted curvature tensor
(\ref{a2rct}) of the $A_2$ isometry.

In 3-dimensional notation the isometry condition
(\ref{b2ic}) is written as
\bea
&\hvGm_\mu=\left( \begin{array}{c} \hA_\mu \\
\hB_\mu  \end{array} \right), \nn\\
&\hD_\mu \hn_1=\hB_\mu \times \hn_2-(\pro_\mu \lambda) \hn_1, \nn\\
&\hD_\mu \hn_2=-\hB_\mu \times \hn_1-(\pro_\mu \lambda) \hn_2.
\eea
From this we have
\bea
&\hA_\mu =A_\mu^1 \hn_1 + A_\mu^2 \hn_2
+(\hn_1 \cdot \pro_\mu \hn_2) \hn_3 \nn\\
&=\Big(\dfrac{e^{\lambda}}{\sqrt2} \Gamma_\mu
+\dfrac{\hn_2\cdot\pro_\mu\hn_3}{2}\Big) \hn_1
-\Big(\dfrac{e^{\lambda}}{\sqrt2} \tGm_\mu
+\dfrac{\hn_1\cdot\pro_\mu\hn_3}{2}\Big) \hn_2 \nn\\
&+(\hn_1 \cdot \pro_\mu \hn_2 )\hn_3,  \nn\\
&\hB_\mu = B_\mu^1 \hn_1 + B_\mu^2 \hn_2 - (\pro_\mu \lambda) \hn_3 \nn\\
&=\Big(\dfrac{e^{\lambda}}{\sqrt2} \tGm_\mu
-\dfrac{\hn_1\cdot\pro_\mu\hn_3}{2}\Big) \hn_1
+\Big(\dfrac{e^{\lambda}}{\sqrt2} \Gamma_\mu
-\dfrac{\hn_2\cdot\pro_\mu\hn_3}{2}\Big) \hn_2  \nn\\
&-(\pro_\mu \lambda)\hn_3,  \nn\\
&A_\mu^1=\dfrac{e^{\lambda}}{\sqrt2}\big(\Gamma_\mu-\tC_\mu^1 \big),
~~A_\mu^2=-\dfrac{e^{\lambda}}{\sqrt2}\big(\tGm_\mu-\tC_\mu^2 \big), \nn\\
&B_\mu^1=\dfrac{e^{\lambda}}{\sqrt2} \big(\tGm_\mu+\tC_\mu^2 \big),
~~B_\mu^2=\dfrac{e^{\lambda}}{\sqrt2}\big(\Gamma_\mu+\tC_\mu^1 \big),  \nn\\
&\tC_\mu^1=-\dfrac{e^{-\lambda}}{\sqrt 2}\hn_2\cdot\pro_\mu\hn_3, \nn\\
&\tC_\mu^2=-\dfrac{e^{-\lambda}}{\sqrt 2} \hn_1 \cdot\pro_\mu \hn_3,
\eea
so that
\bea
&\hA_\mu=-\hn_3 \times \hB_\mu
+\dfrac 12 \epsilon_{ijk}(\hn_i \times \hn_j) \hn_k  \nn\\
&=B_\mu^2 \hn_1-B_\mu^1 \hn_2
+\dfrac 12 \epsilon_{ijk}(\hn_i \times \hn_j) \hn_k,  \nn\\
&\hB_\mu=\hn_3 \times \hA_\mu-\pro_\mu \hn_3-(\pro_\mu \lambda) \hn_3 \nn\\
&=-A_\mu^2 \hn_1 +A_\mu^1 \hn_2-\pro_\mu \hn_3-(\pro_\mu \lambda) \hn_3.
\eea
Notice that both $\hA_\mu$ and $\hB_\mu$ have non-vanishing $\hn_3$
components.

With
\bea
\hvR_\mn=\left( \begin{array}{c} \hA_\mn \\
\hB_\mn  \end{array} \right) \nn
\eea
we have
\bea
&\hA_\mn=\pro_\mu \hA_\nu-\pro_\nu \hA_\mu
+\hA_\mu \times \hA_\nu-\hB_\mu \times \hB_\nu \nn\\
&=\dfrac{e^\lambda}{\sqrt 2}(\Gamma_\mn+H_\mn) \hn_1
-\dfrac{e^\lambda}{\sqrt 2}(\tGm_\mn+\tH_\mn) \hn_2 \nn\\
&=A_\mn^1 \hn_1 + A_\mn^2 \hn_2, \nn\\
&\hB_\mn=\pro_\mu \hB_\nu-\pro_\nu \hB_\mu
+\hA_\mu \times \hB_\nu+\hB_\mu \times \hA_\nu  \nn\\
&=\hD_\mu \hB_\nu-\hD_\nu \hB_\mu \nn\\
&=\dfrac{e^\lambda}{\sqrt 2}(\tGm_\mn+\tH_\mn) \hn_1
+\dfrac{e^\lambda}{\sqrt 2}(\Gamma_\mn+H_\mn) \hn_2  \nn\\
&=B_\mn^1 \hn_1 + B_\mn^2 \hn_2,
\eea
where
\bea
&H_\mn = \pro_\mu \tC_\nu^1 - \pro_\nu \tC_\nu^1
=\dfrac{e^{-\lambda}}{\sqrt2}
\Big(-\hn_1 \cdot(\pro_\mu \hn_1 \times \pro_\nu \hn_1) \nn\\
&+\hn_2 \cdot (\pro_\mu \lambda \pro_\nu \hn_3
-\pro_\nu \lambda \pro_\mu \hn_3) \Big), \nn\\
&\tH_\mn= \pro_\mu \tC_\nu^2 - \pro_\nu \tC_\nu^2
= \dfrac{e^{-\lambda}}{\sqrt2}
\Big(\hn_2 \cdot(\pro_\mu \hn_2 \times \pro_\nu \hn_2) \nn\\
&-\hn_3 \cdot (\pro_\mu \lambda \pro_\nu \hn_1 -\pro_\nu \lambda
\pro_\mu \hn_1) \Big), \nn\\
&A_\mn^1 = B_\mn^2 =\dfrac{e^\lambda}{\sqrt 2}
(\pro_\mu K_\nu - \pro_\nu K_\mu),  \nn\\
&A_\mn^2 = -B_\mn^1 =-\dfrac{e^\lambda}{\sqrt 2} (\pro_\mu
\tK_\nu - \pro_\nu \tK_\mu), \nn\\
&K_\mu=\Gamma_\mu +\tC_\mu^1,
~~~\tK_\mu=\tGm_\mu+\tC_\mu^2,
\eea
so that
\bea
&\hA_\mn=-\hn_3 \times \hB_\mn,
~~~\hB_\mn= \hn_3 \times \hA_\mn.
\eea
Notice that both $\hA_\mn$ and $\hB_\mn$ are orthogonal to $\hn_3$,
although $\hA_\mu$ and $\hB_\mu$ are not.

With this we obtain the full gauge potential of Lorentz group
by adding the valence connection $\vZ_\mu$,
\bea
& \vGm_\mu = \hvGm_\mu + \vZ_\mu, \nn\\
& \vk \cdot \vZ_\mu = \vtk \cdot \vZ_\mu = 0.
\label{b2vc}
\eea
With
\bea
&\vZ_\mu = J_\mu \vk -\tJ_\mu \vtk +L_\mu \vl -\tL_\mu \vtl,  \nn\\
&J_\mu=\vj \cdot \vZ_\mu,~~~\tJ_\mu=\vtj \cdot \vZ_\mu, \nn\\
&L_\mu=\vl \cdot \vZ_\mu,~~~\tL_\mu=\vtl \cdot \vZ_\mu,
\eea
we have
\bea
&\dZ_\mn= \hD_\mu \vZ_\nu-\hD_\nu \vZ_\mu \nn\\
&=U_\mn \vj -\tU_\mn \vtj +(\pro_\mu J_\nu-\pro_\nu J_\mu) \vk
-(\pro_\mu \tJ_\nu-\pro_\nu \tJ_\mu) \vtk  \nn\\
&+ (\cD_\mu L_\nu-\cD_\nu L_\mu) \vl
-(\cD_\mu \tL_\nu-\cD_\nu \tL_\mu) \vtl, \nn\\
&U_\mn = \tK_\mu L_\nu -K_\mu \tL_\nu
-\tK_\nu L_\mu+K_\nu \tL_\mu, \nn\\
&\tU_\mn =K_\mu L_\nu +\tK_\mu \tL_\nu
-K_\nu L_\mu-\tK_\nu \tL_\mu, \nn\\
&\cD_\mu L_\nu = \pro_\mu L_\nu +K_\mu \tJ_\nu-\tK_\mu J_\nu, \nn\\
&\cD_\mu \tL_\nu=\pro_\mu \tL_\nu-K_\mu J_\nu-\tK_\mu \tJ_\nu, \nn\\
&\pZ_\mn=\vZ_\mu \times \vZ_\nu=V_\mn \vk +\tV_\mn \vtk, \nn\\
&V_\mn=J_\mu \tL_\nu +\tJ_\mu L_\nu-J_\nu \tL_\mu -\tJ_\nu L_\mu, \nn\\
&\tV_\mn=\tJ_\mu \tL_\nu-J_\mu L_\nu+J_\nu L_\mu-\tJ_\nu \tL_\mu,
\eea
so that
\bea
&\vZ_\mn=\dZ_\mn+\pZ_\mn =U_\mn \vj -\tU_\mn \vtj \nn\\
&+(\cD_\mu J_\nu-\cD_\nu J_\mu)~\vk
-(\cD_\mu \tJ_\nu-\cD_\nu \tJ_\mu)~\vtk \nn\\
&+(\cD_\mu L_\nu-\cD_\nu L_\mu)~\vl
-(\cD_\mu \tL_\nu-\cD_\nu \tL_\mu)~\vtl, \nn\\
&\cD_\mu J_\nu = \pro_\mu J_\nu-\tL_\mu J_\nu-L_\mu \tJ_\nu, \nn\\
&\cD_\mu \tJ_\nu= \pro_\mu \tJ_\nu-\tL_\mu \tJ_\nu -L_\mu J_\nu.
\eea
Notice that in this case the kinetic part $\dZ_\mn$ contains
all six components, but the potential part $\pZ_\mn$ has only
$\vk$ and $\vtk$ components.
With this we have the full curvature tensor
\bea
&\vR_\mn =\hvR_\mn+\dZ_\mn+\pZ_\mn \nn\\
&=(\Gamma_\mn+H_\mn+U_\mn) \vj
-(\tGm_\mn+\tH_\mn+\tU_\mn) \vtj \nn\\
&+(\cD_\mu J_\nu-\cD_\nu J_\mu) \vk
-(\cD_\mu \tJ_\nu-\cD_\nu \tJ_\mu) \vtk \nn\\
&+(\cD_\mu L_\nu-\cD_\nu L_\mu) \vl
-(\cD_\mu \tL_\nu-\cD_\nu \tL_\mu) \vtl \nn\\
&=(\cD_\mu K_\nu-\cD_\nu K_\mu) \vj
-(\cD_\mu \tK_\nu-\cD_\nu \tK_\mu) \vtj \nn\\
&+(\cD_\mu J_\nu-\cD_\nu J_\mu) \vk
-(\cD_\mu \tJ_\nu-\cD_\nu \tJ_\mu) \vtk \nn\\
&+(\cD_\mu L_\nu-\cD_\nu L_\mu) \vl
-(\cD_\mu \tL_\nu-\cD_\nu \tL_\mu) \vtl \nn\\
&=K_\mn \vj - \tK_\mn \vtj + J_\mn \vk - \tJ_\mn \vtk \nn\\
&+ L_\mn \vl - \tL_\mn \vtl \nn\\
&\cD_\mu K_\nu= \pro_\mu K_\nu +\tL_\mu K_\nu -L_\mu \tK_\nu, \nn\\
&\cD_\mu \tK_\nu= \pro_\mu \tK_\nu -\tL_\mu \tK_\nu -L_\mu K_\nu, \nn\\
&K_\mn =\Gamma_\mn+H_\mn+U_\mn=\cD_\mu K_\nu-\cD_\nu K_\mu, \nn\\
&\tK_\mn =\tGm_\mn+\tH_\mn+\tU_\mn
=\cD_\mu \tK_\nu-\cD_\nu \tK_\mu, \nn\\
&J_\mn = \cD_\mu J_\nu-\cD_\nu J_\mu,
~~~\tJ_\mn= \cD_\mu \tJ_\nu-\cD_\nu \tJ_\mu, \nn\\
&L_\mn = \cD_\mu L_\nu-\cD_\nu L_\mu,  \nn\\
&\tL_\mn= \cD_\mu \tL_\nu-\cD_\nu \tL_\mu,
\label{b2ct1}
\eea
or equivalently
\bea
&R_\mn^{~~ab}=\vR_\mn \cdot \vI^{ab} \nn\\
&= K_\mn j^{ab} - \tK_\mn \tj^{ab} +J_\mn k^{ab} -\tJ_\mn \tk^{ab} \nn\\
&+ L_\mn l^{ab} - \tL_\mn \tl^{ab},
\label{b2ct2}
\eea
This is the $B_2$ decomposition of the curvature tensor.

With complex notation
\bea
&\vk_\pm=\dfrac{1}{\sqrt2}(\vk\pm i\vl),
~~~~\vtk_\pm=\dfrac{1}{\sqrt2}(\vtk\pm i\vtl), \nn\\
&Z_\mu=\dfrac{1}{\sqrt2}(J_\mu+iL_\mu),
~~~~\tZ_\mu= \dfrac{1}{\sqrt2}(\tJ_\mu+i\tL_\mu), \nn\\
&Z_\mu'=\dfrac{1}{\sqrt2}(K_\mu+iL_\mu)
=Z_\mu-\dfrac{1}{\sqrt2} B^-_\mu, \nn\\
&\tZ_\mu'= \dfrac{1}{\sqrt2}(\tK_\mu+i\tL_\mu)
=\tZ_\mu-\dfrac{1}{\sqrt2}\tB^-_\mu,  \nn\\
&B^\pm_\mu=J_\mu \pm K_\mu,~~~\tB^\pm_\mu=\tJ_\mu \pm \tK_\mu,
\eea
we obtain
\bea
&\vZ_\mn= -i(Z'^*_\mu Z'_\nu-Z'^*_\nu Z'_\mu
-\tZ'^*_\mu \tZ'_\nu+\tZ'^*_\nu \tZ'_\mu)\vj \nn\\
&+i(Z'^*_\mu \tZ'_\nu-Z'^*_\nu \tZ'_\mu
-\tZ'^*_\mu Z'_\nu+\tZ'^*_\nu Z'_\mu)\vtj \nn\\
&+(\cD_\mu Z_\nu-\cD_\nu Z_\mu)^*\vk_+ 
-(\cD_\mu \tZ_\nu-\cD_\nu \tZ_\mu)^*\vtk_+, \nn\\
&+(\cD_\mu Z_\nu-\cD_\nu Z_\mu)\vk_- 
-(\cD_\mu \tZ_\nu-\cD_\nu \tZ_\mu)\vtk_-, \nn\\
&\cD_\mu Z_\nu= (\pro_\mu -i\tB^+_\mu)Z_\nu -iB^+_\mu \tZ_\nu^* \nn\\
&+i\tB^-_\mu Z_\nu^*+iB^-_\mu \tZ_\nu.
\eea
With this we have
\bea
&\vR_\mn = (\cD_\mu K_\nu-\cD_\nu K_\mu)~\vj 
-(\cD_\mu \tK_\nu-\cD_\nu \tK_\mu)~\vtj \nn\\
&+(\cD_\mu Z_\nu-\cD_\nu Z_\mu)^*~\vk_+
-(\cD_\mu \tZ_\nu-\cD_\nu \tZ_\mu)^*~\vtk_+  \nn\\
&+(\cD_\mu Z_\nu-\cD_\nu Z_\mu)~\vk_- \nn\\
&-(\cD_\mu \tZ_\nu-\cD_\nu \tZ_\mu)~\vtk_-,
\label{b2ct3}
\eea
or
\bea
&R_\mn^{~~ab}=(\cD_\mu K_\nu-\cD_\nu K_\mu)~j^{ab} 
-(\cD_\mu \tK_\nu-\cD_\nu \tK_\mu)~\tj^{ab}  \nn\\
&+(\cD_\mu Z_\nu-\cD_\nu Z_\mu)^*~k_+^{ab}
-(\cD_\mu \tZ_\nu-\cD_\nu \tZ_\mu)^*~\tk_+^{ab} \nn\\
&+(\cD_\mu Z_\nu-\cD_\nu Z_\mu)~k_-^{ab} \nn\\
&-(\cD_\mu \tZ_\nu-\cD_\nu \tZ_\mu)~\tk_-^{ab}.
\label{b2ct4}
\eea
This should be compared with the $a_2$ result (\ref{a2ct3}) or 
(\ref{a2ct4}).

In 3-dimensional notation, we have
\bea
&\vZ_\mu=\left( \begin{array}{c} \X_\mu \\
\Y_\mu  \end{array} \right), \nn\\
&\X_\mu =\dfrac{e^{-\lambda}}{\sqrt2} \big( J_\mu \hn_1
+\tJ_\mu \hn_2+L_\mu \hn_3 \big),  \nn\\
&\Y_\mu=\dfrac{e^{-\lambda}}{\sqrt2} \big( \tJ_\mu \hn_1
-J_\mu \hn_2+\tL_\mu \hn_3 \big),
\eea
so that
\bea
&\hn_1 \cdot \X_\mu +\hn_2 \cdot \Y_\mu=0,  \nn\\
&\hn_2 \cdot \X_\mu -\hn_1 \cdot \Y_\mu=0,  \nn\\
&\hn \times \Y_\mu= -\hn \times (\hn \times \X_\mu).
\eea
Moreover, with
\bea
&\vZ_\mn=\left( \begin{array}{c} \X_\mn \\
\Y_\mn  \end{array} \right)
=\left( \begin{array}{c} \dX_\mn+\pX_\mn \\
\dY_\mn+\pY_\mn  \end{array} \right),
\eea
we have
\bea
&\dX_\mn=\Big\{\dfrac{e^{\lambda}}{\sqrt2} U_\mn
+\dfrac{e^{-\lambda}}{\sqrt2} (\pro_\mu J_\nu
-\pro_\nu J_\mu) \Big\} \hn_1 \nn\\
&-\Big\{\dfrac{e^{\lambda}}{\sqrt2}\tU_\mn
-\dfrac{e^{-\lambda}}{\sqrt2} (\pro_\mu \tJ_\nu
+\pro_\nu \tJ_\mu) \Big\} \hn_2 +L_\mn \hn_3, \nn\\
&\dY_\mn=\Big\{\dfrac{e^{\lambda}}{\sqrt2}\tU_\mn
+\dfrac{e^{-\lambda}}{\sqrt2} (\pro_\mu \tJ_\nu
-\pro_\nu \tJ_\mu) \Big\}\hn_1 \nn\\
&+\Big\{\dfrac{e^{\lambda}}{\sqrt2} U_\mn
-\dfrac{e^{-\lambda}}{\sqrt2} (\pro_\mu J_\nu
-\pro_\nu J_\mu) \Big\}\hn_2 +\tL_\mn \hn_3, \nn\\
&\pX_\mn=\dfrac{e^{-\lambda}}{\sqrt2} ( V_\mn \hn_1 +\tV_\mn \hn_2), \nn\\
&\pY_\mn=\dfrac{e^{-\lambda}}{\sqrt2} (\tV_\mn \hn_1-V_\mn \hn_2),
\eea
so that
\bea
&\X_\mn=\Big(\dfrac{e^{\lambda}}{\sqrt2}U_\mn
+\dfrac{e^{-\lambda}}{\sqrt2}J_\mn \Big)\hn_1 \nn\\
&-\Big(\dfrac{e^{\lambda}}{\sqrt2}\tU_\mn
-\dfrac{e^{-\lambda}}{\sqrt2} \tJ_\mn \Big)\hn_2 +L_\mn \hn_3, \nn\\
&\Y_\mn=\Big(\dfrac{e^{\lambda}}{\sqrt2}\tU_\mn
+\dfrac{e^{-\lambda}}{\sqrt2} \tJ_\mn \Big)\hn_1 \nn\\
&+\Big(\dfrac{e^{\lambda}}{\sqrt2} U_\mn
-\dfrac{e^{-\lambda}}{\sqrt2} J_\mn \Big)\hn_2 +\tL_\mn \hn_3.
\eea
Finally with
\bea
&\vR_\mn =\left(\begin{array}{c} \A_\mn \\
\B_\mn  \end{array} \right)=\left(\begin{array}{c} \hA_\mn+\X_\mn \\
\hB_\mn+\Y_\mn  \end{array} \right),
\eea
we have
\bea
&\A_\mn=\Big(\dfrac{e^{\lambda}}{\sqrt2}K_\mn
+\dfrac{e^{-\lambda}}{\sqrt2}J_\mn \Big)\hn_1 \nn\\
&-\Big(\dfrac{e^{\lambda}}{\sqrt2}\tK_\mn
-\dfrac{e^{-\lambda}}{\sqrt2} \tJ_\mn \Big)\hn_2 +L_\mn \hn_3, \nn\\
&\B_\mn=\Big(\dfrac{e^{\lambda}}{\sqrt2}\tK_\mn
+\dfrac{e^{-\lambda}}{\sqrt2} \tJ_\mn \Big)\hn_1 \nn\\
&+\Big(\dfrac{e^{\lambda}}{\sqrt2} K_\mn
-\dfrac{e^{-\lambda}}{\sqrt2} J_\mn \Big)\hn_2
+\tL_\mn \hn_3.
\eea
This completes the $B_2$ decomposition of the gravitational
connection.

The above result tells that there exist two different
Abelian decompositions of the gravitational connection
and the curvature tensor which decompose them
into the restricted part and the valence part.
This allows us to decompose the Einstein's
theory in terms of the restricted part
and the valence part.

\section{Decomposition of Einstein's Theory}

Now we are ready to discuss the decomposition of
Einstein's theory. Since the Einstein-Hilbert action is
described by the metric we have to express the above
decomposition of the gravitational
connection in terms of the metric. To do this we use the first order
formalism of Einstein theory. In the absence of the matter field,
the Einstein-Hilbert action in the first order
formalism is given by
\bea
&S[e^\mu_a,~\vGm_\mu]=\dfrac{1}{16\pi G_N} \int \Big(~e~e^\mu_a~e^\nu_b
~\vI^{ab} \cdot \vR_\mn \Big)~d^4x  \nn\\
&=\dfrac{1}{16\pi G_N} \int \Big(\vg_\mn \cdot \vR^\mn\Big)~d^4x, \nn\\
&\vg_\mn = e~ e_\mu^a~e_\nu^b~\vI_{ab},
~~~g_\mn^{~~ab}=e(e_\mu^{~a} e_\nu^{~b}
-e_\nu^{~b} e_\mu^{~a}), \nn\\
&e= {\rm Det}~(e_{\mu a}).
\label{Elag}
\eea
Notice that here we have introduced the Lorentz covariant
four index metric tensor $\vg_\mn$ (which should not be confused with
the two index space-time metric $g_\mn$) which forms
an adjoint representation of Lorentz group.
From (\ref{Elag}) we have the following equation of motion
\bea
&\delta e_{\mu a};~~~\vg_\mn \cdot \vR^{\nu\rho} e_{\rho a}
=R_{\mu a}=0 \nn\\
&\delta \vGm_\mu;~~~D_\mu \vg^\mn= (\pro_\mu+\vGm_\mu \times)\vg^\mn=0,
\label{Eeq}
\eea
where $R_{\mu a}=e^{\nu b} R_{\mn ab}$ is the Ricci tensor.
Clearly the first equation is nothing but the Einstein's
equation in the absence of matter field. But 
the Ricci tensor is written in terms of the gauge potential, 
not the metric. To understand the meaning of the second equation,
notice that the second equation tells that $\vg_\mn$
is invariant under the parallel transport along the
$\pro_\mu$-direction defined by the gauge potential $\vGm_\mu$.
This, of course, puts a strong constraint on the gauge potential.
In fact without much difficulty one can show that 
the second equation uniquely determines $\vGm_\mu$,
\bea
&\vGm_\mu \cdot \vI^{ab}=\Gm_\mu^{~ab} \nn\\
&=\dfrac12 (e^{a\nu} e_{c\mu} \pro^b e^c_{~\nu}
+e^{a\nu}\pro_\mu e^b_{~\nu}+\pro^b e^a_{~\mu} \nn\\
&-e^{b\nu} e_{c\mu} \pro^a e^c_{~\nu}
-e^{b\nu}\pro_\mu e^a_{~\nu}-\pro^a e^b_{~\mu}).
\label{scon}
\eea
This, of course, is the well known equation
of the spin connection $\omega_\mu^{~ab}$,
which confirms that the gauge potential
$\vGm_\mu$ of Lorentz group becomes the spin connection. 
(Here $\Gm_\mu^{~ab}$ becomes the  
torsion-free spin connection, but notice that in general 
it can have torsion when a spinor source is present).
So, the equation $D_\mu \vg^\mn=0$ becomes identical to the
metric-compatibility condition of the connection
\bea
&\nabla_\alpha g_\mn=\pro_\alpha g_\mn
-\Gm_{\alpha\mu}^{~~~\rho} g_{\rho\nu}
-\Gm_{\alpha\nu}^{~~~\rho} g_{\mu\rho}=0, \nn\\
&g_\mn=\eta_{ab} e_\mu^{~a}~e_\nu^{~b},
\eea
which requires the space-time metric $g_\mn$ to be invariant
under the parallel transport defined by the connection.
So in this formalism $D_\mu \vg^\mn=0$ plays the role of
$\nabla_\alpha g_\mn=0$. This confirms that the above
equation (\ref{Eeq}) describes the Einstein's general relativity.

With this preliminary, we discuss the decomposition of
Einstein's theory with the $A_2$ isometry 
(the space-like isometry) first.
For this we introduce two projection operators which project out
the isometry components,
\bea
&\vSi_{ab}=l_{ab}~\vl-\tl_{ab}~\vtl, \nn\\
& \vP_{ab}=\vI_{ab}-\vSi_{ab}=l^1_{ab}~\vl_1-\tl^1_{ab}~\vtl_1
+l^2_{ab}~\vl_2-\tl^2_{ab}~\vtl_2, \nn\\
&\Si_{ab}^{~~cd}=l_{ab} l^{cd}-\tl_{ab} \tl^{cd}, \nn\\
&\Pi_{ab}^{~~cd}=I_{ab}^{~~cd}-\Si_{ab}^{~~cd},  \nn\\
&\vZ_\mu \cdot \vSi_{ab}=0,~~~\vZ_\mu \cdot \vP_{ab}=Z_\mu^{~ab}.
\eea
Clearly $\vSi_{ab}$ and $\vP_{ab}$ become projection operators
in the sense that
\bea
&\vSi_{ab} \cdot \vSi^{cd} =\dfrac 12 \Si_{ab}^{~~mn}\Si_{mn}^{~~cd}
=\Si_{ab}^{~~cd}, \nn\\
&\vP_{ab} \cdot \vP^{cd}=\dfrac 12 \Pi_{ab}^{~~mn}\Pi_{mn}^{~~cd}
=\Pi_{ab}^{~~cd},  \nn\\
&\vSi_{ab} \cdot \vP^{cd}=0.
\eea
Now we can express the Einstein-Hilbert action as
\bea
&S[e^\mu_a,~\Gm_\mu,~\tGm_\mu,~\vZ_\mu]
=\dfrac{1}{16\pi G_N} \int \Big\{~\vg_\mn \cdot \vR^\mn \nn \\
&+\lambda (\vl^2-1) + \tilde \lambda (\vl \cdot \vtl)
+\lambda_\mu (\vl \cdot \vZ^\mu)
+\tilde \lambda_\mu (\vtl \cdot \vZ^\mu)\Big\}~d^4x, \nn\\
&\vR_\mn =\hvR_\mn+(\hD_\mu \vZ_\nu-\hD_\nu \vZ_\mu)
+\vZ_\mu \times \vZ_\nu  \nn\\
&=(\cD_\mu L_\nu-\cD_\nu L_\mu) \vl
-(\cD_\mu \tL_\nu-\cD_\nu \tL_\mu) \vtl \nn\\
&+ (\hD_\mu \vZ_\nu-\hD_\nu \vZ_\mu),
\label{a2lag1}
\eea
where $\lambda's$ are the Lagrange
multipliers. From this we
get the following equations of motion
\bea
&\delta e_{\mu c};~~(e~e_\mu^a~e_\nu^b)
\big[(\cD^\nu L^\rho-\cD^\rho L^\nu)~l_{ab} \nn\\
&-(\cD^\nu \tL^\rho-\cD^\rho \tL^\nu)\tl_{ab}
+(\hD^\nu \vZ^\rho-\hD^\rho \vZ^\nu) \cdot \vP_{ab} \big] e_{\rho c} \nn\\
&=0, \nn\\
&\delta \Gamma_\nu;~~ \pro_\mu (e~e_a^\mu ~e_b^\nu ~l^{ab})
+\vl  \cdot ( \vZ_\mu \times \vg^\mn)=0,  \nn\\
& \delta \tGm_\nu;~~\pro_\mu (e~e_a^\mu~e_b^\nu~\tl^{ab})
+\vtl  \cdot ( \vZ_\mu \times \vg^\mn)=0,  \nn\\
& \delta \vZ_\nu;~~ \hD_\mu (e~e_a^\mu ~e_b^\nu ~\vP^{ab})
+(e~e_a^\mu e_b^\nu) \big[(\vZ_\mu \times \vl)l^{ab} \nn\\
&-\vZ_\mu \times \vtl) \tl^{ab} \big]=0.
\label{a2Eeq1}
\eea
Notice that, using the fact that $\hD_\mu \vl=\hD_\mu \vtl=0$, we can
combine the last three equations into a single equation,
\bea
D_\mu \vg^\mn=0.
\eea
But this is precisely the second equation of (\ref{Eeq}),
which confirms that (\ref{a2Eeq1}) is equivalent to (\ref{Eeq}).

To clarify the meaning of the above equation
we define the restricted metric $\hvg_\mn$ decomposing
$\vg_\mn$
\bea
&\vg_\mn = \hvg_\mn + \vG_\mn,  \nn\\
&\hvg_\mn =e~e_\mu^a~e_\nu^b~\vSi_{ab}=G_\mn ~\vl -\tG_\mn \vtl, \nn\\
&\vG_\mn= e~e_\mu^a~e_\nu^b~\vP_{ab}=G_\mn^1 \vl_1-\tG_\mn^1 \vtl_1  \nn\\
&+G_\mn^2 \vl_2-\tG_\mn^2 \vtl_2, \nn\\
&G_\mn=e~e_\mu^a~e_\nu^b~l_{ab},
~~~\tG_\mn =e~ e_\mu^a~e_\nu^b~\tl_{ab}, \nn\\
&G_\mn^1=e~e_\mu^a~e_\nu^b~l_{ab}^1,
~~~\tG_\mn^1 =e~ e_\mu^a~e_\nu^b~\tl_{ab}^1, \nn\\
&G_\mn^2=e~e_\mu^a~e_\nu^b~l_{ab}^2,
~~~\tG_\mn^2 =e~ e_\mu^a~e_\nu^b~\tl_{ab}^2.
\label{gdef}
\eea
Notice that
\bea
&\tG_\mn = \dfrac 12 \epsilon_{abcd} e~e_\mu^{a} ~e_\nu^{b}
~l^{cd}
=\dfrac e2 \epsilon_{\mn cd} l^{cd} \nn\\
&= \dfrac 12 \epsilon_{\mn\rho\sigma} G^{\rho\sigma} = G_\mn^d,  \nn\\
&\tG_\mn^1 = G_\mn^{1~d},~~~~~\tG_\mn^2 = G_\mn^{2~d}.
\eea
Clearly the two two-forms $G_\mn$ and $\tG_\mn$ can be viewed
to represent the restricted metric which are dual to each other.
With this (\ref{a2Eeq1}) has the following compact expression
\bea
&G_\mn (\cD^\nu L^\rho-\cD^\rho L^\nu)
-\tG_\mn (\cD^\nu \tL^\rho-\cD^\rho \tL^\nu) \nn\\
&+\vG_\mn \cdot (\hD^\nu \vZ^\rho-\hD^\rho \vZ^\nu)=0, \nn\\
& \pro_\mu G^\mn +\vl \cdot (\vZ_\mu \times \vG^\mn) =0, \nn\\
& \pro_\mu \tG^\mn +\vtl \cdot (\vZ_\mu \times \vG^\mn) =0, \nn\\
& \hD_\mu \vG^\mn +\vZ_\mu \times (G^\mn ~\vl -\tG^\mn \vtl)=0,
\label{a2Eeq2}
\eea
or equivalently
\bea
&G_\mn (\cD^\nu L^\rho-\cD^\rho L^\nu)
-\tG_\mn (\cD^\nu \tL^\rho-\cD^\rho \tL^\nu) \nn\\
&=-G_\mn^i (\cD^\nu Z^\rho_i -\cD^\rho Z^\nu_i)
+\tG_\mn^i (\cD^\nu \tZ^\rho_i-\cD^\rho \tZ^\nu_i), \nn\\
&\pro_\mu G^\mn=-\epsilon_{ij} (Z^i_\mu G_j^\mn-\tZ^i_\mu \tG_j^\mn), \nn\\
&\pro_\mu \tG^\mn=-\epsilon_{ij} (Z^i_\mu \tG_j^\mn+\tZ^i_\mu G_j^\mn), \nn\\
& \pro_\mu G_i^\mn =\epsilon_{ij}(L_\mu G_j^\mn -\tL_\mu
\tG_j^\mn - Z^j_\mu G^\mn +\tZ^i_\mu \tG^\mn), \nn\\
& \pro_\mu \tG_i^\mn =\epsilon_{ij}(L_\mu \tG_j^\mn +\tL_\mu
G_j^\mn -Z^i_\mu \tG^\mn - \tZ^j_\mu G^\mn). \nn\\
&(i,j=1,2,~~~\epsilon_{12}=-\epsilon_{21}=1)
\label{a2Eeq3}
\eea
This suggests that the valence connection $\vZ_\mu$
plays the role of the gravitational source of
the restricted metric.

In 3-dimensional notation
\bea
&\hvg_\mn=\left( \begin{array}{c} \hm_\mn \\
\he_\mn  \end{array} \right),
~~~\vG_\mn=\left( \begin{array}{c} \M_\mn \\
\E_\mn  \end{array} \right), \nn\\
&\vg_\mn=\left( \begin{array}{c} \m_\mn \\
\e_\mn  \end{array} \right)
=\left( \begin{array}{c} \hm_\mn+\M_\mn \\
\he_\mn+\E_\mn  \end{array} \right),
\eea
we have
\bea
&\hm_\mn = G_\mn \hn,~~~~ \he_\mn = \tG_\mn \hn,  \nn\\
&\M_\mn = G^1_\mn \hn_1+G^2_\mn \hn_2,  \nn\\
&\E_\mn = \tG^1_\mn \hn_1+\tG^2_\mn \hn_2,
\eea
so that the Einstein-Hilbert action (\ref{a2lag1}) acquires
the following form
\bea
&S[e^\mu_a,~L_\mu,~\tL_\mu,~Z^i_\mu,~\tZ^i_\mu] \nn\\
&= \dfrac{1}{16\pi G_N} \int \Big\{G_\mn (\cD^\mu L^\nu-\cD^\nu L^\mu) \nn\\
&-\tG_\mn (\cD^\mu \tL^\nu-\cD^\nu \tL^\mu)
+G_\mn^i (\cD^\mu Z^{\nu i} -\cD^\nu Z^{\mu i})  \nn\\
&-\tG_\mn^i (\cD^\mu \tZ^{\nu i}-\cD^\nu \tZ^{\mu i}) \Big\} d^4x.
\label{a2lag2}
\eea
From this we can reproduce (\ref{a2Eeq3}).
This completes the $A_2$ decomposition (the space-like 
decomposition) of Einstein's theory.

We can repeat the same procedure with the $B_2$ isometry 
(the light-like isometry) to obtain
the desired decomposition of Einstein's equation.
With the Einstein-Hilbert action
\bea
&S[e^\mu_a,~\Gm_\mu,~\tGm_\mu,~\vZ_\mu]
=\dfrac{1}{16\pi G_N} \int
\Big\{\vg_\mn \cdot~\vR_\mn  \nn \\
&+\lambda~\vj^2 +\tilde \lambda (\vj \cdot \vtj) 
+\lambda_\mu (\vk \cdot \vZ^\mu)
+\tilde \lambda_\mu (\vtk \cdot \vZ^\mu)\Big\}~d^4x,\nn\\
&\vR_\mn =\hvR_\mn+(\hD_\mu \vZ_\nu-\hD_\nu \vZ_\mu)
+\vZ_\mu \times \vZ_\nu  \nn\\
&=(\cD_\mu K_\nu-\cD_\nu K_\mu)~\vj
-(\cD_\mu \tK_\nu-\cD_\nu \tK_\mu)~\vtj  \nn\\
&+(\cD_\mu J_\nu-\cD_\nu J_\mu) \vk
-(\cD_\mu \tJ_\nu-\cD_\nu \tJ_\mu) \vtk \nn\\
&+(\cD_\mu L_\nu-\cD_\nu L_\mu) \vl
-(\cD_\mu \tL_\nu-\cD_\nu \tL_\mu) \vtl,
\label{b2lag1}
\eea
we get following equations of motion
\bea
& \delta e_{\mu c} ~;~~(e~e^a_\mu ~e^b_\nu)
\Big( (\cD^\nu K^\rho-\cD^\rho K^\nu)~j_{ab} \nn\\
&-(\cD^\nu \tK^\rho-\cD^\rho \tK^\nu)~\tj_{ab}
+\vZ^{\nu\rho} \cdot \vP_{ab} \big) e_{\rho c}=0, \nn\\
&\delta \Gamma_\nu ~;~~ \pro_\mu (e~e_a^\mu ~e_b^\nu ~j^{ab})
+\vj \cdot (\vZ_\mu \times \vg^\mn)=0 \nn\\
&\delta \tGm_\nu ~;~~ \pro_\mu (e~e_a^\mu ~e_b^\nu ~\tj^{ab})
+\vtj \cdot (\vZ_\mu \times \vg^\mn)=0 \nn\\
&\delta \vZ_\nu ~;~~ \hD_\mu (e~e_a^\mu ~e_b^\nu ~\vP^{ab})
+(e~e_a^\mu ~e_b^\nu) \big[ (\vZ_\mu \times \vk)j^{ab} \nn\\
&-(\vZ_\mu\times \vtk)\tj^{ab} \big]=0,
\label{b2Eeq1}
\eea
where now
\bea
&\vP_{ab}=k_{ab}~\vj-\tk_{ab}~\vtj +l_{ab}~\vl-\tl_{ab}~\vtl
=\vI_{ab}-\vSi_{ab}, \nn\\
&\vSi_{ab}=j_{ab}~\vk-\tj_{ab}~\vtk.
\eea
Notice that here $\vP_{ab}$ and $\vSi_{ab}$ do not make
projection operators, because
\bea
&\vP_{ab} \cdot \vSi^{cd}=k_{ab}~j^{cd}-\tk_{ab}~\tj^{cd} \neq 0.
\eea
Now, again we can combine the last three equations of 
(\ref{b2Eeq1}) into a single equation with the help of 
$\hD_\mu \vj=\hD_\mu \vtj=0~$,
\bea
D_\mu \vg^\mn=0. \nn
\eea
This confirms that (\ref{b2Eeq1}) is equivalent to (\ref{Eeq}),
which tells that (\ref{b2lag1}) describes the Einstein's gravity.

Now, with
\bea
&\vg_\mn = \hvg_\mn +\vG_\mn, \nn\\
&\hvg_\mn=e~e_\mu^a~e_\nu^b~\vSi^{ab}=\cJ_\mn~\vk-\ctJ_\mn~\vtk, \nn\\
&\vG_\mn= e~e_\mu^a~e_\nu^b~\vP^{ab}  \nn\\
&=\cK_\mn~\vj-\ctK_\mn~\vtj +\cL_\mn~\vl-\ctL_\mn~\vtl,  \nn\\
&\cJ_\mn=e~e_\mu^a~e_\nu^b~j_{ab},
~~~\ctJ_\mn=e~e_\mu^a~e_\nu^b~\tj_{ab}, \nn\\
&\cK_\mn=e~e_\mu^a~e_\nu^b~k_{ab},
~~~\ctK_\mn=e~e_\mu^a~e_\nu^b~\tk_{ab}, \nn\\
&\cL_\mn=e~e_\mu^a~e_\nu^b~l_{ab},
~~~\ctL_\mn=e~e_\mu^a~e_\nu^b~\tl_{ab},
\eea
the equation (\ref{b2Eeq1}) is written as
\bea
&\cJ_\mn (\cD^\nu K^\rho-\cD^\rho K^\nu)
-\ctJ_\mn (\cD^\nu \tK^\rho-\cD^\rho \tK^\nu) \nn\\
&+\vG_\mn \cdot \vZ^{\nu\rho} =0, \nn\\
& \pro_\mu \cJ^\mn +\vj \cdot (\vZ_\mu \times \vG^\mn)=0, \nn\\
& \pro_\mu \ctJ^\mn +\vtj \cdot (\vZ_\mu \times \vG^\mn)=0, \nn\\
& \hD_\mu \vG^\mn +\vZ_\mu \times (\cJ_\mn~\vk-\ctJ_\mn~\vtk)=0,
\label{b2Eeq2}
\eea
or equivalently
\bea
&\cJ_\mn (\cD^\nu K^\rho-\cD^\rho K^\nu)
-\ctJ_\mn (\cD^\nu \tK^\rho-\cD^\rho \tK^\nu) \nn\\
&=-\cK_\mn (\cD^\nu J^\rho-\cD^\rho J^\nu)
+\ctK_\mn (\cD^\nu \tJ^\rho-\cD^\rho \tJ^\nu) \nn\\
&-\cL_\mn (\cD^\nu L^\rho-\cD^\rho L^\nu)
+\ctL_\mn (\cD^\nu \tL^\rho-\cD^\rho \tL^\nu), \nn\\
& \pro_\mu \cJ^\mn = L_\mu \ctJ^\mn +\tL_\mu \cJ^\mn
-J_\mu \ctL^\mn -\tJ_\mu \cL^\mn, \nn\\
& \pro_\mu \ctJ^\mn = -L_\mu \cJ^\mn +\tL_\mu \ctJ^\mn
+J_\mu \cL^\mn -\tJ_\mu \ctL^\mn, \nn\\
& \pro_\mu \cK^\mn = -L_\mu \ctK^\mn -\tL_\mu \cK^\mn
-K_\mu \cL^\mn +\tK_\mu \ctL^\mn, \nn\\
& \pro_\mu \ctK^\mn = L_\mu \cK^\mn -\tL_\mu \ctK^\mn
+K_\mu \ctL^\mn +\tK_\mu \cL^\mn, \nn\\
& \pro_\mu \cL^\mn = K_\mu \cJ^\mn -\tK_\mu \ctJ^\mn
+J_\mu \ctK^\mn +\tJ_\mu \cK^\mn , \nn\\
& \pro_\mu \ctL^\mn = -K_\mu \ctJ^\mn-\tK_\mu \cJ^\mn \nn\\
&-J_\mu \cK^\mn +\tJ_\mu \ctK^\mn.
\label{b2Eeq3}
\eea
Remember that $\cJ_\mn,~\cK_\mn,~\cL_\mn$ and 
$\ctJ_\mn,~\ctK_\mn,~\ctL_\mn$ are dual
to each other. Here again the valence connection becomes
the gravitational source of the restricted metric.

In 3-dimensional notation we have
\bea
&\vg_\mn=\left( \begin{array}{c} \m_\mn \\
\e_\mn  \end{array} \right)
=\left( \begin{array}{c} \hm_\mn+\M_\mn \\
\he_\mn+\E_\mn  \end{array} \right), \nn\\
&\hm_\mn = \dfrac{e^{-\lambda}}{\sqrt2} (\cJ_\mn \hn_1+\ctJ_\mn \hn_2), \nn\\
&\he_\mn = \dfrac{e^{-\lambda}}{\sqrt2} (\ctJ_\mn \hn_1-\cJ_\mn \hn_2), \nn\\
&\hm_\mn = \n_3 \times \he_\mn,
~~~\he_\mn=-\n_3 \times \hm_\mn,  \nn\\
&\M_\mn = \dfrac{e^{\lambda}}{\sqrt2} (\cK_\mn \hn_1+\ctK_\mn \hn_2)
+\cL_\mn \hn_3, \nn\\
&\E_\mn = \dfrac{e^{\lambda}}{\sqrt2} (-\ctK_\mn \hn_1+ \cK_\mn
\hn_2)+\ctL_\mn \hn_3,
\eea
so that the Einstein-Hilbert action (\ref{b2lag1})
is expressed as
\bea
&S[e^\mu_a,~K_\mu,~\tK_\mu,~J_\mu,~\tJ_\mu,~L_\mu,~\tL_\mu] \nn\\
&= \dfrac{1}{16\pi G_N} \int
\Big\{ \cJ_\mn (\cD^\mu K^\nu-\cD^\nu K^\mu)  \nn\\
&-\ctJ_\mn (\cD^\mu \tK^\nu-\cD^\nu \tK^\mu)
+\cK_\mn (\cD^\mu J^\nu-\cD^\nu J^\mu) \nn\\
&-\ctK_\mn(\cD^\mu \tJ^\nu-\cD^\nu \tJ^\mu)
+\cL_\mn (\cD^\mu L^\nu-\cD^\nu L^\mu) \nn\\
&-\ctL_\mn (\cD^\mu \tL^\nu-\cD^\nu \tL^\mu) \Big\}~d^4x.
\label{b2lag2}
\eea
From this we can reproduce (\ref{b2Eeq3}).
This completes the $B_2$ decomposition 
(the light-like decomposition) of Einstein's theory.

The above decompositions of Einstein's theory are
interesting in their own right. They allow us a new interpretation
of Einstein's theory which sheds a new light on
gravitation. More importantly they have far reaching consequences,
as we will see \cite{grg0}.

\section{Discussions}

 In this paper we have discussed the Abelian decomposition of
Einstein's theory. Imposing proper magnetic isometries
to the gravitational connection, we have shown how to decompose the
gravitational connection and the curvature tensor into the
restricted part of the maximal Abelian subgroup $H$ of Lorentz group
$G$ and the valence part of $G/H$ component 
which plays the role of the Lorentz
covariant gravitational source of the restricted connection, without
compromising the general invariance. 
This tells that the Einstein's theory can be viewed as a
theory of the restricted gravity made of the restricted connection
in which the valence connection plays the role of the gravitational
source of the restricted gravity. 
We show that there are two
different Abelian decompositions of Einstein's theory,
the space-like $A_2$ decomposition and the light-like 
$B_2$ decomposition,
because Lorentz group has two maximal Abelian
subgroups.

A common difficulty in non-Abelian
gauge theory and gravitation is the highly non-linear self
interaction. In gauge theory one can simplify this non-linear
interaction by separating the gauge covariant valence part from the
Abelian part of the potential. This simplification
has played a crucial role to establish the Abelian dominance 
in non-Abelian gauge theory. Here we have shown that we can 
simplify the gravitational interaction 
exactly the same way, treating Einstein's
theory as a gauge theory of Lorentz group.

An important ingredient of the decomposition is the concept of
Lorentz covariant four-index metric tensor $\vg_\mn$ which
replaces the role of the two-index space-time metric $g_\mn$.
We have shown that the metric-compatibility condition
of the connection $\nabla_\alpha g_\mn=0$ is replaced by
the gauge covariant condition $D_\mu \vg^\mn=0$.

From theoretical point of view, the above
decomposition of gravitation differs from the Abelian
decomposition of non-Abelian gauge theory in one important respect.
In gauge theory the fundamental ingredient is the gauge potential,
and the decomposition of the potential provides
a complete decomposition of the theory. On the other hand 
in gravitation the fundamental field is assumed to be 
the metric, not the potential (the connection), but 
our decomposition is based on the connection. 
So one might wonder whether one can have an Abelian 
decomposition of Einstein's theory based on the metric. 
Of course our decomposition allows us to have the
the decomposition of the metric, but only indirectly
through the equation of motion, as we have seen.
It would be very interesting to see 
if one can actually decompose the metric explicitly, 
and decompose Einstein's theory in terms of the metric.

This touches a subtle but very important issue on the essence 
of gravitation. The issue here is what describes gravity. 
In Einstein's theory gravity is described by 
the metric alone. But, as Cartan has pointed out long time ago,
torsion could also be an important part of gravity \cite{cartan}.
And indeed in the gauge formalism of Einstein's theory 
the connection (the gauge potential of Lorentz group) naturally 
includes torsion \cite{prd76b,hehl}. This tells that our Abelian 
decomposition is a decomposition of a {\it generalized} 
Einstein's theory which includes torsion. This is very important, 
because torsion could generate new topological structures 
which are absent in Einstein's gravity.   

Clearly the above decomposition of (generalized) Einstein's theory 
has many interesting implications. First of all, this tells
that one can actually construct a restricted theory of gravitation
made of the restricted part of connection alone
which has the full Lorentz invariance,
or equivalently the full general invariance, by excluding the
Lorentz covariant valence part of connection from the theory.
In other words, one can have a theory of gravitation which
actually is much simpler than Einstein's theory,
which nevertheless has the full general invariance
and thus retains all topological characteristics of 
Einstein's theory. This is very important, because
this tells that the restricted gravity describes the core
dynamics of Einstein's theory. This point can play 
a crucial role for us to understand the quantum gravity. 

Furthermore, the decomposition makes the topology of
generalized Einstein's theory more transparent as well as 
interesting. Indeed with the Abelian decomposition 
we can study the topological structures of the theory 
more easily, because the topological characteristics are 
imprinted in the magnetic symmetry. 
For example, the $A_2$ decomposition makes it clear that 
the topology of Einstein's theory is closely related to
the topology of $SU(2)$ gauge theory. This is natural,
because $SU(2)$ forms a subgroup of Lorentz group.
This strongly implies that gravity
may allow the gravito-magnetic monopole which has 
the monopole topology $\pi_2(S^2)$ \cite{cho91,new}. 

Perhaps more importantly, this strongly implies that 
classical space-time in generalized Einstein's theory
may have the multiple vacua similar to 
what we find in $SU(2)$ gauge theory. This turns out to be true. 
In fact with a proper magnetic isometry we can construct all possible 
vacuum space-times, and show that they are classified by 
exactly the same multiple vacua that we have
in $SU(2)$ gauge theory which can be classified by 
the knot topology $\pi_3(S^3)=\pi_3(S^2)$ \cite{grg0}.
This has a far reaching consequence.
Just as in $SU(2)$ gauge theory, the multiple vacua
in gravity can be unstable against quantum fluctuation. 
And there is a real possibility that gravitational interaction
may allow the gravito-instantons which can connect 
topologically distinct vacua and thus allow the
vacuum tunnelling. Clearly this will have an important 
implication in quantum gravity.
Of course the ``gravitational istantons"
which have finite action in Euclidian space-time have been 
discussed before, because it was hoped that
these Euclidian configurations could play an important role
in quantum gravity \cite{hawk,egu}. But so far 
it has not been known exactly what role
they play in quantum gravity. In any case they have not been
associated with the above multiple vacuum space-time.  

The above discussions raise more questions 
need to be answered. A first question is the existence of 
a real gravito-instanton which can actually make the 
quantum tunnelling between topologically different vacuum 
space-times. It would be intersting to find
such gravito-instanton (if exist), and see whether any of 
the known finite action Euclidian 
configurations can make the desired quantum tunnelling.
Another question is the role (if any) of torsion in this
multiple vacuua and vacuum tunneling. Again it would be 
important to know whether Einstein's theory 
in the absence of torsion can admit multiple vacuua and/or 
vacuum tunneling. If this is so, it would be very interesting 
to construct the multiple vacuua and gravito-instanton 
in terms of metric. The details of the subject 
and related issues will be discussed 
separately in an accompanying paper \cite{grg3}.

{\bf ACKNOWLEDGEMENT}

~~~The work is supported in part by the BSR Program 
(Grant KRF-2007-314-C00055) of Korea Research Foundation 
and in part by the ABRL Program (Grant R14-2003-012-01002-0) 
and the International Cooperation Program of Korea Science 
and Engineering Foundation.

\end{document}